\documentclass[aps,pre,reprint,groupedaddress,longbibliography,showpacs]{revtex4-1}

\usepackage{amsmath,amssymb,epsfig,bbm,MnSymbol,verbatim}

\begin{document}


\title{Periodic Thermodynamics of Open Quantum Systems}


\author{Kay Brandner\textsuperscript{1}}
\author{Udo Seifert\textsuperscript{2}}
\affiliation{\textsuperscript{1}Department of Applied Physics, Aalto University, 
00076 Aalto, Finland\\
\textsuperscript{2}II. Institut f\"ur Theoretische Physik, Universit\"at Stuttgart, 70550 Stuttgart, Germany}


\date{\today}

\begin{abstract}
The thermodynamics of quantum systems coupled to periodically
modulated heat baths and work reservoirs is developed.
By identifying affinities and fluxes, the first and second law
are formulated consistently. 
In the linear response regime, entropy production becomes a quadratic
form in the affinities.
Specializing to Lindblad-dynamics, we identify the corresponding 
kinetic coefficients in terms of correlation functions of the 
unperturbed dynamics.
Reciprocity relations follow from symmetries with respect to time 
reversal.
The kinetic coefficients can be split into a classical and a 
quantum contribution subject to a new constraint, which follows 
from a natural detailed balance condition.
This constraint implies universal bounds on efficiency and
power of quantum heat engines. 
In particular, we show that Carnot efficiency can not be reached
whenever quantum coherence  effects are present, i.e., when the 
Hamiltonian used for work extraction does not commute with the bare
system Hamiltonian. 
For illustration, we specialize our universal results to a driven 
two-level system in contact with a heat bath of sinusoidally modulated
temperature.  
\end{abstract}


\maketitle

\newcommand{\R}{\mathbb{R}}
\newcommand{\Tc}{T^{{{\rm c}}}}
\newcommand{\Th}{T^{{{\rm h}}}}
\newcommand{\T}{\mathcal{T}}
\renewcommand{\L}{\mathsf{L}}
\newcommand{\D}{\mathsf{D}}
\renewcommand{\H}{\mathsf{H}}
\newcommand{\K}{\tilde{\L}^{0\dagger}}
\newcommand{\X}{\mathsf{X}}
\newcommand{\Y}{\mathsf{Y}}
\newcommand{\rc}{\varrho^{{{\rm c}}}}
\newcommand{\req}{\varrho^{{{\rm eq}}}}
\newcommand{\rad}{\varrho^{{{\rm ad}}}}
\newcommand{\kb}{k_{{{\rm B}}}}
\newcommand{\F}{\mathcal{F}}
\newcommand{\J}{\mathcal{J}}
\newcommand{\TR}{\mathsf{T}}
\newcommand{\Li}{L^{{{\rm ins}}}}
\newcommand{\Lr}{L^{{{\rm ret}}}}
\newcommand{\x}{\mathbf{x}}
\newcommand{\y}{\mathbf{y}}
\newcommand{\z}{\mathbf{z}}
\newcommand{\hc}{\eta_{{{\rm C}}}}
\newcommand{\Jc}{J_q^{{{\rm c}}}}
\newcommand{\ve}{\varepsilon}
\newcommand{\ta}{a}
\newcommand{\tb}{b}
\newcommand{\g}{\mathbf{g}}
\newcommand{\G}{\mathbf{G}}
\newcommand{\W}{\mathbb{W}^{0t}}

\newcommand{\lint}{\int_0^1\!\!\! d\lambda\;}
\newcommand{\tauint}{\int_0^\infty\!\!\! d\tau\;}
\renewcommand{\tint}{\int_0^\T\!\!\!  dt \;}
\newcommand{\ttauint}{\int_0^\T\!\!\! dt
\int_0^\infty\!\!\! d\tau \;}

\newcommand{\ed}[2]{\Bigl\llangle #1,#2\Bigr\rrangle}
\newcommand{\eds}[2]{\left\llangle #1,#2\right\rrangle}
\newcommand{\ev}[2]{\Bigl\langle #1,#2\Bigr\rangle}
\newcommand{\evs}[2]{\left\langle #1,#2\right\rangle}
\newcommand{\tr}[1]{{{\rm tr}}\left\{#1\right\}}
\newcommand{\pro}[1]{\left|#1\right\rangle\left\langle #1 \right|}

\allowdisplaybreaks

\section{Introduction}

In a thermodynamic cycle, a working fluid is driven by a sequence of
control operations, e.g., compressions and expansions through a moving
piston, and temperature variations such that its initial state is
restored after one period \cite{Callen1985}. 
The net effect of such a process thus consists in the transfer of 
heat and work between a set of controllers and reservoirs external to
the system. 
This concept was originally designed to link the operation principle
of macroscopic machines such as Otto or Diesel engines with the 
fundamental laws of thermodynamics. 
As a paramount result, these efforts inter alia unveiled that the
efficiency of any heat engine operating between two reservoirs of 
respectively constant temperature is bounded by the Carnot value.

\begin{figure}[h!]
\centering
\epsfig{file=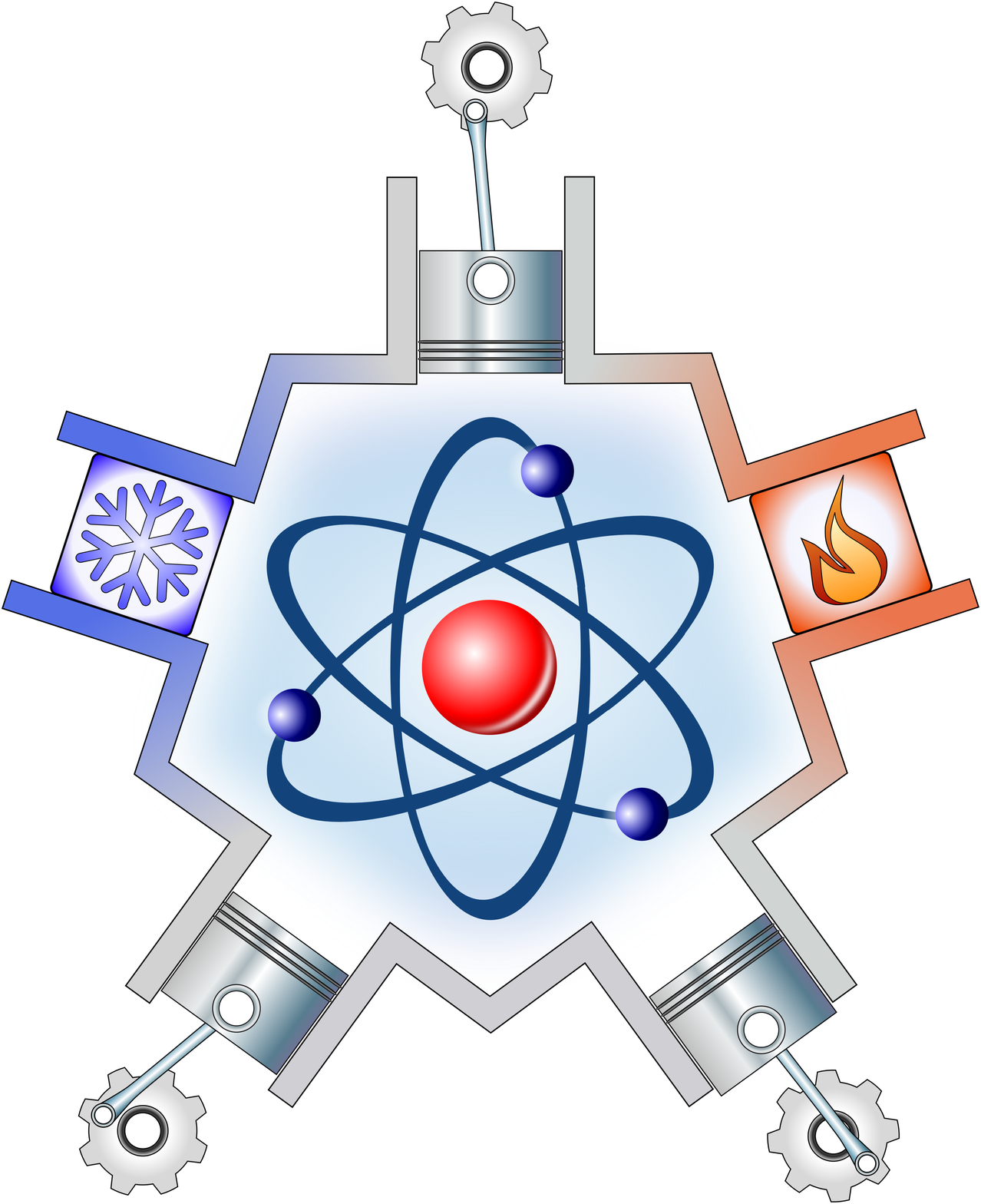,scale=0.074}
\caption{Illustration of a periodically driven open quantum system.
The energy of the system, symbolically shown as an atom confined in a
chamber, is modulated by three external controllers, each of which is
represented by a reciprocating piston. 
Simultaneously, heat is exchanged with one cold and one hot reservoir.
\label{Fig_Engine}}
\end{figure}

During the last decade, thermodynamic cycles have been implemented on
increasingly smaller scales. 
Particular landmarks of this development are mesoscopic heat engines,
whose working substance consists of a single colloidal particle
\cite{Blickle2011,Martinez2015} or a micrometer-sized mechanical
spring \cite{Steeneken2010}. 
Recently, a further milestone was achieved by crossing the border to 
the quantum realm in experiments realizing cyclic thermodynamic 
processes with objects like single electrons \cite{Koski2014,
Pekola2015} or atoms \cite{Abah2012,Roßnagel2015}. 
In the light of this progress, the question emerges whether quantum
effects might allow to overcome classical limitations such as the 
Carnot bound \cite{Gardas2015b}. 
Indeed, there is quite some evidence that the performance of thermal
devices can, in principle, be enhanced by exploiting, for example,
coherence effects \cite{Scully2010,Scully2011,Horowitz2014,
Brandner2015d,Mitchison2015a,Uzdin2015,Hofer2015a}, non-classical 
reservoirs \cite{Scully2003,Dillenschneider2009,Rossnagel2014,
Abah2014,Manzano2015} or the properties of superconducting materials
\cite{Hofer2015}. 
These studies are, however, mainly restricted to specific models and
did so far not reveal a universal mechanism that would allow cyclic 
energy converters to benefit from quantum phenomena. 

The theoretical description of quantum thermodynamic cycles faces two
major challenges. 
First, the external control parameters are typically varied  
non-adiabatically. 
Therefore, the state of the working fluid can not be described by an
instantaneous Gibbs-Boltzmann distribution, an assumption 
inherent to conventional macroscopic thermodynamics. 
Second, the degrees of freedom of the working substance are inevitably
affected by both, thermal and quantum fluctuations, which must be
consistently taken into account.

In this paper, we take a first step towards a general framework
overcoming both of these obstacles. 
To this end, we consider the generic setup of Fig.~\ref{Fig_Engine},
i.e., a small quantum system, which is weakly coupled to a set of
thermal reservoirs with periodically time-dependent temperature and
driven by multiple controllers altering its Hamiltonian. 
Building on the scheme originally proposed in \cite{Brandner2015f},
we develop a universal approach that describes the corresponding 
thermodynamic process in terms of time-independent affinities and 
cycle-averaged fluxes. 
Focusing on mean values thereby allows us to avoid subtleties 
associated with the definitions of heat and work for single 
realizations \cite{Horowitz2012,Horowitz2013a,Horowitz2014b,
Jarzynski2015,Hanggi2015}.
In borrowing a term first coined by Kohn \cite{Kohn2001} in the
context of quantum systems interacting with strong laser-fields, we
refer to this theory as periodic thermodynamics of open quantum 
systems. 

In the linear response regime, where temperature and energy variations
can be treated perturbatively, a quantum thermodynamic cycle is 
fully determined by a set of time-independent kinetic coefficients. 
Such quantities were first considered in \cite{Izumida2009,
Izumida2010,Izumida2015a} for some specific models of Brownian heat
engines and later obtained on a more general level for classical 
stochastic systems with continuous \cite{Brandner2015f,Bauer2016} and
discrete states \cite{Proesmans2015,Proesmans2015a}.
Here, we prove two universal properties of the quantum kinetic
coefficients for open systems following a Markovian time evolution. 
First, we derive a generalized reciprocity relation stemming  from 
microreversibility. 
Second, we establish a whole hierarchy of constraints, which 
explicitly account for coherences between unperturbed energy 
eigenstates and lie beyond the laws of classical thermodynamics. 

For quantum heat engines operating under linear response conditions,
these relations imply strong restrictions showing that quantum
coherence is generally detrimental to both, power and efficiency. 
In particular, the Carnot bound can be reached only if the external
driving protocol commutes with the unperturbed Hamiltonian of the
working substance, which then effectively behaves like a discrete
classical system. 
As one of our key results, we can thus conclude that any thermal 
engine, whose performance is truly enhanced through quantum
effects, must be equipped with components that are not covered by our
general setup as for example non-equilibrium reservoirs or feedback 
mechanisms. 

The rest of the paper is structured as follows. 
We begin with introducing our general framework in 
Sec.~\ref{SecFramework}. 
In Sec.~\ref{SecMarkovDyn} we outline a set of requirements on the
Lindblad generator, which  ensure the thermodynamic consistency of the
corresponding time-evolution. 
Using this dynamics we then focus on quantum kinetic coefficients in
Sec.~\ref{SecGKC}. 
Sec.~\ref{SecQTM} is devoted to the derivation of general bounds on 
the figures of performance of quantum heat engines. 
We work out an explicit example for such a device in Sec.\ref{SecEx}. 
Finally, we conclude in Sec.~\ref{SecCon}. 

\section{Framework}\label{SecFramework}

\subsection{General Scheme}

As illustrated in Fig. \ref{Fig_Engine}, we consider an open quantum
system, which is mechanically driven by $N_w$ external controllers and
attached to $N_q$ heat baths with respectively time-dependent
temperature $T_\nu(t)$.
The total Hamiltonian of the system is given by
\begin{equation}\label{GTDFullHamiltonian}
H(t)\equiv H^0 + \sum_{j=1}^{N_w} \Delta_j H \; g_{wj}(t),
\end{equation}
where $H^0$ corresponds to the free Hamiltonian, the dimensionless
operator $g_{wj}(t)$ represents the driving exerted by the controller
$j$ and the scalar energy $\Delta_j H$ quantifies the strength of this
perturbation.  
For this set-up, the first law reads
\begin{equation}\label{GTDFistLaw}
\dot{U}(t) = \sum_{\nu=1}^{N_q} \dot{Q}_\nu(t)
           - \sum_{j=1}^{N_w} \dot{W}_j(t)
\end{equation}
with dots indicating derivatives with respect to time throughout the
paper. 
By expressing the internal energy
\begin{equation}\label{GTDIntEnergy}
U(t)\equiv \tr{H(t)\varrho(t)}
\end{equation}
in terms of the density matrix $\varrho(t)$, which characterizes the
state of the system, we obtain the power extracted by the controller 
$j$ \cite{Alicki1979b,Kosloff1984a,Geva1994},
\begin{equation}\label{GTDPower}
\dot{W}_j(t)\equiv -\Delta_j H \; \tr{\dot{g}_{wj}(t)\varrho(t)}.
\end{equation}
Furthermore, the total heat current absorbed from the environment
becomes
\begin{equation}\label{GTDHeatCurr}
\sum_{\nu=1}^{N_q} \dot{Q}_\nu(t) \equiv \tr{H(t)\dot{\varrho}(t)},
\end{equation}
where $\tr{\bullet}$ denotes the trace operation from 
\eqref{GTDIntEnergy} onwards. 
We note that \eqref{GTDIntEnergy} does not lead to a microscopic 
expression for the individual heat current $\dot{Q}_\nu(t)$ related to
the reservoir $\nu$. 
This indeterminacy arises because thermal perturbations can not be
included in the total Hamiltonian $H(t)$. 
Taking them into account explicitly rather requires to specify the 
mechanism of energy exchange between system and each reservoir. 

Still, any dissipative dynamics must be consistent with the second
law, which requires 
\begin{equation}\label{GTDEntropyProd}
\dot{S}(t)\equiv \dot{S}_{{{\rm sys}}}(t)
-\sum_{\nu=1}^{N_q}\frac{\dot{Q}_{\nu}(t)}{T_\nu(t)}\geq 0,
\end{equation}
with $\dot{S}(t)$ denoting the total rate of entropy production. 
The first contribution showing up here corresponds to the change in 
the von Neumann-entropy of the system
\begin{equation}\label{GTDVonNeumann}
S_{{{\rm sys}}}(t)\equiv - \kb\tr{\varrho(t)\ln\varrho(t)},
\end{equation}
where $\kb$ denotes Boltzmann's constant.
The second one accounts for the entropy production in the environment. 
We now focus on the situation, where the Hamiltonian $H(t)$ and the 
temperatures $T_\nu(t)$ are $\T$-periodic in time.
After a certain relaxation time, the density matrix of the system will
then settle to a periodic limit cycle $\rc(t)=\rc(t+\T)$. 
Consequently, after averaging over one period, \eqref{GTDEntropyProd}
becomes
\begin{equation}\label{GTDAvEntropy}
\dot{S}\equiv\frac{1}{\T}\tint \dot{S}(t) 
= -\sum_{\nu=1}^{N_q}\tint \frac{\dot{Q}_\nu(t)}{T_\nu(t)},
\end{equation}
i.e., no net entropy is produced in the system during a full operation
cycle. 

The entropy production in the environment can be attributed to the
individual controllers and reservoirs by parametrizing the 
time-dependent temperatures as  
\cite{Brandner2015f}
\begin{equation}\label{GTDTemp}
T_\nu(t)\equiv \frac{\Th_\nu\Tc}{\Th_\nu+(\Tc-\Th_\nu)
\gamma_{q\nu}(t)}.
\end{equation}
Here, $\Tc\leq T_\nu(t)$ denotes the reference temperature, $\Th_\nu$
is the maximum temperature reached by the reservoir $\nu$ and the $0
\leq\gamma_{q\nu}(t)\leq 1$ are dimensionless functions of time.
Inserting \eqref{GTDFistLaw}, \eqref{GTDPower} and \eqref{GTDTemp} 
into \eqref{GTDAvEntropy} yields
\begin{equation}\label{GTDBilinEntropy}
\dot{S}= \sum_{j=1}^{N_w}\F_{wj} J_{wj} 
       + \sum_{\nu=1}^{N_q} \F_{q\nu} J_{q\nu}
\end{equation}
with generalized affinities 
\begin{equation}
\F_{wj}\equiv\frac{\Delta_j H}{\Tc}, \qquad
\F_{q\nu}\equiv \frac{1}{\Tc}-\frac{1}{\Th_\nu}
\end{equation}
and generalized fluxes 
\begin{align}
J_{wj} &\equiv\frac{1}{\T}\tint \tr{\dot{g}_{wj}(t)\rc(t)},
\label{GTDWorkFlux}\\
J_{q\nu} &\equiv\frac{1}{\T}\tint \gamma_{q\nu}(t) \dot{Q}_\nu(t).
\label{GTDHeatFlux}
\end{align}
Expression \eqref{GTDBilinEntropy}, which constitutes our first main
result, resembles the generic form of the total rate of entropy 
production known from conventional irreversible thermodynamics
\cite{Callen1985}. 
It shows that the mean entropy, which must be generated to maintain
a periodic limit cycle in an open quantum system, can be expressed as
a bilinear form of properly chosen fluxes and affinities. 
Each pair thereby corresponds to a certain source of mechanical 
or thermal driving. 

\subsection{Linear Response Regime}

A particular advantage of our approach is that it allows a systematic
analysis of the linear-response regime, which is defined by the
temporal gradients $\Delta_\nu T\equiv \Th_\nu-\Tc$ and 
$\Delta_j H$ being small compared to their respective reference 
values $\Tc$ and 
\begin{equation}
E^{{{\rm eq}}}\equiv \tr{H^0\req}.
\end{equation}
Here, 
\begin{equation}\label{GTDEqDist}
\req\equiv\exp[-H^0/(k_B \Tc)]/Z^0
\end{equation}
denotes the equilibrium state of the system and $Z^0$ the canonical 
partition function. 

The generalized fluxes \eqref{GTDWorkFlux} and \eqref{GTDHeatFlux}
then become
\begin{equation}\label{GTDLinRespRel}
J_\alpha \equiv \sum_\beta L_{\alpha\beta}\F_\beta
+\mathcal{O}\left(\Delta^2\right), 
\end{equation}
where
\begin{equation}\label{GTDLinAff}
\F_{wj}=\frac{\Delta_j H}{\Tc}
\quad\text{and}\quad 
\F_{q\nu}= \frac{\Delta_\nu T}{(\Tc)^2}
+ \mathcal{O}\left(\Delta^2\right).
\end{equation}
The combined indices $\alpha,\beta\equiv wj,q\nu$ allow a compact
notation. 
The generalized kinetic coefficients $L_{\alpha\beta}$ introduced in 
\eqref{GTDLinRespRel} are conveniently arranged in a matrix 
\begin{equation}
\mathbb{L}\equiv\left(\!\begin{array}{cc}
\mathbb{L}_{ww} & \mathbb{L}_{wq}\\
\mathbb{L}_{qw} & \mathbb{L}_{qq}
\end{array}\!\right)
\end{equation}
with 
\begin{equation}\label{GTDDefBlockMat}
\mathbb{L}_{AB}\equiv \left(\!\begin{array}{ccc}
L_{A1, B1} & \cdots & L_{A1,BN_B}\\
\vdots     & \ddots & \vdots\\
L_{AN_A,B1} & \cdots & L_{AN_A,BN_B}
\end{array}\!\right)
\quad (A,B\equiv w,q).
\end{equation}
Inserting \eqref{GTDLinRespRel} into \eqref{GTDBilinEntropy} shows 
that, in the linear response regime, the mean entropy production per
operation cycle becomes
\begin{equation}\label{GTDDefSymOM}
\dot{S}=\sum_{\alpha\beta} L_{\alpha\beta}\F_\alpha\F_\beta
=\frac{
\boldsymbol{\F}^t\left(\mathbb{L}+\mathbb{L}^t\right)\boldsymbol{\F}}
{2}
\equiv \boldsymbol{\F}^t\mathbb{L}^{{{\rm s}}}\boldsymbol{\F}
\end{equation}
with $\boldsymbol{\F}\equiv (\F_{w1},\dots,\F_{wN_w},\F_{q1},\dots
\F_{qN_q})^t$. 
Consequently, the second law $\dot{S}\geq 0$ implies that the
symmetric part $\mathbb{L}^{{{\rm s}}}$ of the matrix $\mathbb{L}$
must be positive semi-definite.

\section{Markovian Dynamics}\label{SecMarkovDyn}
So far, we have introduced a universal framework for the thermodynamic
description of periodically driven open quantum systems.
We will now apply this scheme to systems, whose time-evolution is
governed by the Markovian quantum master equation 
\cite{Breuer2006}
\begin{equation}\label{MDLindblad}
\partial_t\varrho(t) = \L(t)\varrho(t)
\end{equation}
with generator
\begin{equation}\label{MDGenerator}
\L(t)\equiv\H(t) + \sum_{\nu=1}^{N_q}\D_\nu(t).
\end{equation}
Here, the super-operator 
\begin{equation}\label{MDUnitary}
\H(t)\bullet\equiv
-\frac{i}{\hbar}\left[H(t),\bullet\right]
\end{equation}
describes the unitary dynamics of the bare system, where 
$[\bullet,\circ]$ indicates the usual commutator and $\hbar$ denotes
Planck's constant. 
The influence of the reservoir $\nu$ is taken into account by the 
dissipation super-operator
\begin{equation}\label{MDDissipator}
\D_\nu(t)\bullet\equiv
\sum_{\sigma}\frac{\Gamma_{\nu}^{\sigma}(t)}{2}
\left(
[V_{\nu}^{\sigma}(t)\bullet ,V_{\nu}^{\sigma\dagger}(t)]
+ [V^{\sigma}_{\nu}(t),\bullet V_{\nu}^{\sigma\dagger}(t)]
\right)
\end{equation}
with time-dependent rates $\Gamma^\sigma_\nu(t)\geq 0$ and 
Lindblad-operators $V^\sigma_\nu(t)$.
As a consequence of this structure, the time-evolution generated by 
\eqref{MDLindblad} can be shown to preserve trace and complete
positivity of the density matrix  $\varrho(t)$ \cite{Rivas,Breuer2015}.
Furthermore, after a certain relaxation time, it leads to a periodic
limit cycle $\varrho^{{{\rm c}}}(t)=\varrho^{{{\rm c}}}(t+\T)$ for any 
initial condition \cite{Kosloff2013}. 
For later purpose, we introduce here also the unperturbed generator 
\begin{align}
\left.\L(t)\right|_{\F=0} & \equiv
\L^0\equiv\H^0 +\sum_{\nu=1}^{N_q}\D^0_\nu \quad\text{with}\nonumber\\
\H^0\bullet & \equiv -\frac{i}{\hbar}[H^0,\bullet]
\quad\text{and} \nonumber\\
\D^0_\nu\bullet & \equiv \sum_\sigma \frac{\Gamma^{\sigma}_\nu}{2}
\left([V^\sigma_\nu\bullet,V^{\sigma\dagger}_\nu]
     +[V^\sigma_\nu,\bullet V^{\sigma\dagger}_\nu]\right),
\label{MDUnpertLindbladF}
\end{align}
where we assume the set of free Lindblad-operators $\{V^\sigma_\nu\}$ 
to be self-adjoint and irreducible
\footnote{A set of operators $\mathcal{A}\equiv\{A_k\}$ is 
self-adjoint if for any $A_k\in\mathcal{A}$ also $A_k^\dagger\in 
\mathcal{A}$. 
The set is irreducible if the only operators commuting with all
elements of $\mathcal{A}$ are scalar multiples of the identity.}.

The structure \eqref{MDGenerator} of the generator $\L(t)$ naturally
leads to microscopic expressions for the individual heat currents 
$\dot{Q}_\nu(t)$.
Specifically, after insertion of \eqref{MDLindblad} and
\eqref{MDGenerator}, the total heat uptake \eqref{GTDHeatCurr}
can be written in the form
\begin{equation}
\sum_{\nu=1}^{N_q}\dot{Q}_\nu(t) 
= \sum_{\nu=1}^{N_q}\tr{H(t)\D_\nu(t)\varrho(t)},
\end{equation}
which suggests the definition \cite{,Spohn1978a,Alicki1979b,Geva1994}

\begin{equation}\label{MDDetailedHeatCurrent}
\dot{Q}_\nu(t)\equiv\tr{H(t)\D_\nu(t)\varrho(t)}.
\end{equation}
This identification has been shown to be consistent with the second
law \eqref{GTDEntropyProd} if the dissipation super-operators 
$\D_\nu(t)$ fulfil \cite{Alicki1979b,Spohn1978}
\begin{equation}\label{MDAdiabaticInv}
\D_\nu(t)\varrho^{{{\rm ins}}}_\nu(t)=0, 
\end{equation}
where 
\begin{equation}
\varrho^{{{\rm ins}}}_\nu(t)\equiv \exp[-H(t)/(\kb T_\nu(t))]/Z_\nu(t)
\end{equation}
with $Z_\nu(t)\equiv\tr{\exp[-H(t)/(\kb T_\nu(t))]}$ denotes an
instantaneous equilibrium state. 
In appendix \ref{SecSecondLaw}, we show that, if the reservoirs are 
considered as mutually independent, \eqref{MDAdiabaticInv} is also a
necessary condition for \eqref{GTDEntropyProd} to hold. 

After specifying the dissipative dynamics of the system, the 
expressions for the generalized fluxes \eqref{GTDWorkFlux} and 
\eqref{GTDHeatFlux} can be made more explicit. 
First, integrating by parts with respect to $t$ in 
\eqref{GTDWorkFlux} and then eliminating $\dot{\varrho}^{{{\rm c}}}(t)$
using \eqref{MDLindblad} yields 
\begin{equation}\label{MDWorkFlux}
J_{wj} = -\frac{1}{\T}\tint \tr{g_{wj}(t)\L(t)\rc (t)}.
\end{equation}
The corresponding boundary terms vanish, since $g_{wj}(t)$ and 
$\rc (t)$ are $\T$-periodic in $t$. 
Second, by plugging \eqref{MDDetailedHeatCurrent} into 
\eqref{GTDHeatFlux}, we obtain the microscopic expression
\begin{equation}\label{MDHeatFlux}
J_{q\nu}=\frac{1}{\T}\tint \gamma_{q\nu}(t)
\tr{H(t)\D_\nu(t)\rc(t)}
\end{equation}
for the generalized heat flux extracted from the reservoir $\nu$. 

As a second criterion for thermodynamic consistency, we require that
the unperturbed dissipation super-operators $\D^0_\nu$  fulfill the
quantum detailed balance relation \cite{Alicki1976,
Kossakowski1977,Frigerio1984,Majewski1984}
\begin{equation}\label{MDDetailedBalanceLoc}
\D_\nu^0\req=\req\D^{0\dagger}_\nu.
\end{equation}
This condition ensures that, in equilibrium, the net rate of
transitions between each individual pair of unperturbed energy
eigenstates is zero.  
Note that, in \eqref{MDAdiabaticInv}, $\D_\nu(t)$ acts on the operator
exponential, while \eqref{MDDetailedBalanceLoc} must be read as an 
identity between super-operators. 
Furthermore, throughout this paper, the adjoint of super-operators is
indicated by a dagger and understood with respect to the 
Hilbert-Schmidt scalar product \cite{Breuer2006}, i.e., for example 
\begin{equation}
\D^{0\dagger}_\nu\bullet\equiv\sum_\sigma\frac{\Gamma_\nu^\sigma}{2}
\left(V^{\sigma\dagger}_\nu[\bullet,V^\sigma_\nu]
+[V^{\sigma\dagger}_\nu,\bullet]V^\sigma_\nu\right).
\end{equation}

For systems, which can be described on a finite-dimensional Hilbert 
space, \eqref{MDDetailedBalanceLoc} implies that the super-operator
$\D_\nu^0$ can be written in the natural form \cite{Alicki1976,
Kossakowski1977,Frigerio1984}
\begin{align}
& \D_\nu^0\bullet = \frac{1}{2}\sum_{\sigma} 
\Gamma_\nu^{\sigma}
\left([V^{\sigma\dagger}_\nu\bullet,V^\sigma_\nu]+
     [V^{\sigma\dagger}_\nu,\bullet V^\sigma_\nu]\right)\nonumber\\
&\hspace*{2.5cm}
     + \bar{\Gamma}^\sigma_\nu
\left([V^\sigma_\nu\bullet,V^{\sigma\dagger}_\nu]+
     [V^\sigma_\nu,\bullet V^{\sigma\dagger}_\nu]\right)
\quad\text{with}\nonumber\\[9pt]
& \bar{\Gamma}^\sigma_\nu\equiv
\Gamma^\sigma_\nu\exp[-\varepsilon^\sigma_\nu/(\kb\Tc)],
\quad \Gamma^\sigma_\nu>0, \nonumber\\[9pt]
& [H^0,V^\sigma_\nu] =
\varepsilon^\sigma_\nu V^\sigma_\nu, 
\quad\text{and}\quad \varepsilon^\sigma_\nu\geq 0 . 
\label{ApxGKCPropUnpert1}
\end{align}
Conversely, however, these conditions imply 
\eqref{MDDetailedBalanceLoc} even if the dimension of the underlying
Hilbert space is infinite. 
Therefore, the results of the subsequent sections, which rely on both,
\eqref{MDDetailedBalanceLoc} and \eqref{ApxGKCPropUnpert1}, are not 
restricted to systems with a finite spectrum. 
They rather apply whenever the unperturbed dissipation super-operators
$\D^0_\nu$ have the form \eqref{ApxGKCPropUnpert1} as, for example, in
the standard description of the dissipative harmonic oscillator 
\cite{Kosloff1984,Breuer2006,Horowitz2012}.

The characteristics of the generator $\L(t)$ discussed in this section
form the basis for our subsequent analysis. 
Although they are justified by phenomenological arguments involving
the second law and the principle of microreversibility, it is worth
noting that most of these properties can be derived from first 
principles. 
Specifically, \eqref{MDDetailedBalanceLoc} and 
\eqref{ApxGKCPropUnpert1} have been shown to emerge naturally from
a general microscopic model for a time-independent
open system in the weak-coupling limit \cite{Davies1976,
Carmichael1976,Kossakowski1977,Spohn1978a,Davies1978a}.
Moreover, for a single reservoir of constant temperature, the 
time-dependent relation \eqref{MDAdiabaticInv} has been derived
using a similar method under the additional assumption that the 
time-evolution of the bare driven system is slow on the time-scale of
the reservoirs \cite{Davies1978,Albash2012}.
In the opposite limit of fast driving, this microscopic approach can
be combined with Floquet theory to obtain an essentially different 
type of Lindblad-generator \cite{Zerbe1995,Breuer1997,Kohler1998,
Szczygielski2013,Kosloff2013,Cuetara2015a}.
The thermodynamic interpretation of the corresponding time-evolution
is, however, not yet settled. 
The question how a thermodynamically consistent master equation
for a general set-up involving a driven system, multiple reservoirs and 
time-dependent temperatures can be derived from first principles is 
still open at this point.

\section{Generalized Kinetic Coefficients}\label{SecGKC}

\subsection{Microscopic Expressions}

Solving the master equation \eqref{MDLindblad} within a first order
perturbation theory and exploiting the properties of the generator
$\L(t)$ discussed in the previous section leads to explicit 
expressions  for the generalized kinetic coefficients 
\eqref{GTDLinRespRel}. 
For convenience, we relegate this procedure to the first part of 
appendix \ref{ApxGKC} and present here only the result
\begin{widetext}
\begin{alignat}{3}
L_{wj,wk} & \equiv \Li_{wj,wk} + \Lr_{wj,wk} &&\equiv 
-\frac{1}{\kb\T}\tint\ev{g_{wj}(t)}{\K g_{wk}(t)}
-\frac{1}{\kb\T}\ttauint\ev{g_{wj}(t)}
{\K e^{\K\tau} \K g_{wk}(t-\tau)}, \nonumber\\
L_{wj,q\nu} & \equiv \Li_{wj,q\nu} + \Lr_{wj,q\nu} &&\equiv 
-\frac{1}{\kb\T}\tint\ev{g_{wj}(t)}{\D^{0\dagger}_\nu g_{q\nu}(t)}
-\frac{1}{\kb\T}\ttauint\ev{g_{wj}(t)}
{\K e^{\K\tau} \D^{0\dagger}_\nu g_{q\nu}(t-\tau)}, \nonumber\\
L_{q\nu,wj} & \equiv \Li_{q\nu,wj}+ \Lr_{q\nu,wj} &&\equiv 
-\frac{1}{\kb\T}\tint\ev{g_{q\nu}(t)}{\D^{0\dagger}_\nu g_{wj}(t)}
-\frac{1}{\kb\T}\ttauint\ev{g_{q\nu}(t)}
{\D^{0\dagger}_\nu e^{\K\tau} \K g_{wj}(t-\tau)}, \nonumber\\
L_{q\nu,q\mu} & \equiv \Li_{q\nu,q\mu} + \Lr_{q\nu,q\mu}  &&\equiv 
-\frac{\delta_{\nu\mu}}{\kb\T}\tint\ev{g_{q\nu}(t)}
{\D^{0\dagger}_\nu g_{q\nu}(t)}
-\frac{1}{\kb\T}\ttauint\ev{g_{q\nu}(t)}
{\D^{0\dagger}_\nu e^{\K\tau} \D^{0\dagger}_\mu g_{q\mu}(t-\tau)},
\label{GKC}
\end{alignat}
\end{widetext}
where $\delta_{\nu\mu}$ denotes the Kronecker symbol, $g_{wj}(t)$ was 
defined in \eqref{GTDFullHamiltonian}, 
\begin{equation}\label{GKCDefgq}
g_{q\nu}(t)\equiv -\gamma_{q\nu}(t)H^0
\end{equation}
and 
\begin{equation}\label{GKCDefLTR}
\K\equiv \H^0 + \sum_{\nu=1}^{N_q}\D^{0\dagger}_\nu.
\end{equation}
Furthermore, we introduced the scalar product \cite{Kubo1998}
\begin{align}\label{GKCScalarProd}
\langle\bullet,\circ\rangle &\equiv\lint \tr{\bullet^\dagger
R^\lambda\circ R^{-\lambda}\req} \quad\text{with}\nonumber\\
R&\equiv\exp[-H^0/(\kb\Tc)]
\end{align}
in the space of operators. 

The two parts of the coefficients $L_{\alpha\beta}$ showing up in 
\eqref{GKC} can be interpreted as follows. 
First, the modulation of the Hamiltonian and the temperatures of the
reservoirs leads to non-vanishing generalized fluxes $J_{wj}$ and
$J_{q\nu}$ even before the system has time to adapt to these 
perturbations.
This effect is captured by the instantaneous coefficients
$\Li_{\alpha\beta}$. 
Second, in responding to the external driving, the state of the system
deviates from thermal equilibrium thus giving rise to the retarded 
coefficients $\Lr_{\alpha\beta}$.
We note that the expressions \eqref{GKC} do not involve the full 
generator $\L(t)$ but only the unperturbed super-operators $\D^0_\nu$ 
and $\H^0$. 
This observation confirms the general principle that linear response
coefficients are fully determined by the free dynamics of the system
and the small perturbations disturbing it \cite{Kubo1998}. 

Compared to the kinetic coefficients recently obtained for 
periodically driven classical systems \cite{Brandner2015f,
Proesmans2015a,Proesmans2015}, the expressions \eqref{GKC} are 
substantially more involved. 
This additional complexity is, however, not due to quantum effects
but rather stems from the presence of multiple reservoirs, which has
not been considered in the previous studies. 
Indeed, as we show in the second part of appendix \ref{ApxGKC}, if 
only a single reservoir is attached to the system, \eqref{GKC}
simplifies to 
\begin{multline}\label{GKCSimplified}
L_{\ta\tb}\equiv  L^{{{\rm ad}}}_{\ta\tb} 
+ L^{{{\rm dyn}}}_{\ta\tb}
=-\frac{1}{\kb\T}\tint \ev{\delta\dot{g}_{\ta}(t)}
{\delta g_{\tb} (t)}\\
 +\frac{1}{\kb\T}\ttauint \ev{\delta\dot{g}_{\ta}(t)}
{e^{\K\tau}\delta\dot{g}_{\tb}(t-\tau)},
\end{multline}
where $\ta,\tb=wj,q1$.
The deviations of the external perturbations from equilibrium are 
thereby defined as 
\begin{equation}\label{GKCEqFluct}
\delta g_{\ta}(t)\equiv g_{\ta}(t)-\tr{g_{a}(t)\req}
= g_{\ta}(t)-\evs{\mathbbm{1}}{g_{\ta}(t)},
\end{equation}
where dots indicate derivatives with respect to $t$ and $\mathbbm{1}$ 
denotes the unity operator. 
Expression \eqref{GKCSimplified} has precisely the same structure
as its classical analogue with the only difference that the scalar
product had to be modified according to \eqref{GKCScalarProd} in oder
to account for the non-commuting nature of quantum observables.

As in the classical case, the single-reservoir coefficients 
\eqref{GKCSimplified} can be split into an adiabatic part $L^{{{\rm
ad}}}_{\ta\tb}$, which persists even for infinitely slow
driving, and a dynamical one $L^{{{\rm dyn}}}_{\ta\tb}$ 
containing finite-time corrections. 
This partitioning, which was originally suggested in 
\cite{Brandner2015f}, is, however, not equivalent to the division
into instantaneous and retarded contributions introduced here. 
In fact, the later scheme is more general than the former one, which
can not be applied when the system is coupled to more than one 
reservoir.
In such set-ups, temperature gradients between distinct reservoirs
typically prevent the existence of a universal adiabatic state, which,
in the case of a single reservoir, is given by the instantaneous
Boltzmann distribution \cite{Proesmans2015}. 

\subsection{Reciprocity Relations}

After deriving the explicit expressions for the generalized kinetic 
coefficients \eqref{GKC}, we will now explore the interrelations
between these quantities. 
To this end, we first have to discuss the principle of microscopic 
reversibility or $T$-symmetry \cite{VanKampen1957,VanKampen1957a,
Agarwal1973}.
A closed and autonomous, i.e., undriven, quantum system is said to
be $T$-symmetric if its Hamiltonian commutes with the anti-unitary
time-reversal operator $T$ \cite{Mazenko2006}. 
In generalizing this concept, here we call an open, autonomous 
system $T$-symmetric if the generator $\L^0$ governing its 
time-evolution fulfills 
\begin{equation}\label{MDDetailedBalance}
\L^0\req \TR = \TR\req\L^{0\dagger},
\end{equation}
where
\begin{equation}
\TR\bullet\equiv T\bullet T^{-1}
\end{equation}
and $\req$ is the stationary state associated with $\L^0$. 
This definition is motivated by the fact that, within the 
weak-coupling approach, \eqref{MDDetailedBalance} arises from the 
$T$-symmetry of the total system including the reservoirs and their
coupling to the system proper \cite{Carmichael1976}. 
Note that, here, we assume the absence of external magnetic fields. 

The condition \eqref{MDDetailedBalance} was first derived by Agarwal
in order to extend the classical notion of detailed balance to the
quantum realm \cite{Agarwal1973}. 
In the same spirit, Kossakowski obtained the relation
\eqref{MDDetailedBalanceLoc} and the structure 
\eqref{ApxGKCPropUnpert1} without reference to time-reversal symmetry. 
Provided that $\L^0$ has the Lindblad form \eqref{MDUnpertLindbladF}, 
the condition \eqref{MDDetailedBalanceLoc} is indeed less restrictive
than \eqref{MDDetailedBalance}. 
In fact, \eqref{MDDetailedBalance} follows from
\eqref{ApxGKCPropUnpert1} and \eqref{MDUnpertLindbladF} under the
additional requirement that \cite{Alicki1976}
\begin{equation}\label{MDHamiltInv}
TH^0 = H^0T \quad\text{and}\quad T V^\sigma_\nu = V^\sigma_\nu T.
\end{equation}

Microreversibility implies an important property of the generalized
kinetic coefficients \eqref{GKC}.
Specifically, if the free Hamiltonian $H^0$ and the free 
Lindblad operators $V^\sigma_\nu$ defined in \eqref{MDUnpertLindbladF}
satisfy \eqref{MDHamiltInv}, i.e., if the unperturbed system is 
$T$-symmetric, we have the reciprocity relations
\begin{equation}\label{GKCRR}
L_{\alpha\beta} [g_\alpha(t),g_\beta(t)] 
=L_{\beta\alpha} [\TR g_\alpha(-t),\TR g_\beta (-t)].
\end{equation}
Here, the $L_{\alpha\beta}$ are regarded as functionals of the
perturbations $g_\alpha (t)$. 
The symmetry \eqref{GKCRR}, which we prove in appendix \ref{ApxRR},
constitutes the analogue of the well-established Onsager-relations
\cite{Onsager1931,Onsager1931a} for periodically driven open quantum
systems. 
Its classical counterpart was recently derived in
\cite{Brandner2015f} for a single reservoir and one external
controller. 
Extensions to classical set-ups with multiple controllers were 
subsequently obtained in \cite{Proesmans2015,Proesmans2015a}.

The quantities $g_{q\nu}(t)$ defined in \eqref{GKCDefgq} are invariant
under the action $\TR$ by virtue of \eqref{MDHamiltInv}. 
Thus, if the modulations of the Hamiltonian fulfill $\TR g_{wj}(t)
= g_{wj}(t)$, \eqref{GKCRR} reduces to 
\begin{equation}\label{GKCRRSym}
L_{\alpha\beta} [g_\alpha(t),g_\beta(t)] 
=L_{\beta\alpha} [g_\alpha(-t),g_\beta (-t)].
\end{equation} 
Furthermore, if the $g_{wj}(t)$ can be written in the form 
\begin{equation}\label{GKCFactCond}
g_{wj}(t) = \gamma_{wj}(t) g_{wj},
\end{equation}
where $\gamma_{wj}(t)\in\mathbb{R}$ and $\TR g_{wj} = g_{wj}$, the 
special symmetry 
\begin{equation}\label{GKCRRFact} 
L_{\alpha\beta} [\gamma_\alpha(t),\gamma_\beta(t)]
= L_{\beta\alpha} [\gamma_\beta(t), \gamma_\alpha(t)]
\end{equation}
holds, which, in contrast to \eqref{GKCRR} and \eqref{GKCRRSym}, does
not involve the reversed protocols (see appendix \ref{ApxRR} for
details).  

\subsection{Quantum Effects}\label{SecClassSys}

We will now explore to what extend the kinetic coefficients 
\eqref{GKC} show signatures of quantum coherence. 
To this end, we assume for simplicity that the spectrum of the 
unperturbed Hamiltonian $H^0$ is non-degenerate. 
A quasi-classical system is then defined by the condition
\begin{equation}\label{GKCCommH}
[H^0,g_{wj}(t)] = 0 \quad\text{for}\quad j=1,\dots,N_w,
\end{equation}
which entails that, up to second-order corrections in $\Delta_j H$
and $\Delta_\nu T$, the periodic state $\rc(t)$ is diagonal in the 
joint eigenbasis of $H^0$ and the perturbations $g_{wj}(t)$ at any 
time $t$.
Thus, the corresponding kinetic coefficients effectively describe a
discrete classical system with periodically modulated energy levels
given by the eigenvalues of the full Hamiltonian $H(t)$. 
This result, which is ultimately a consequence of the detailed balance
structure \eqref{ApxGKCPropUnpert1}, is proven in the first part
of appendix~\ref{ApxQCS}, where we also provide explicit expressions
for the quasi-classical kinetic coefficients 
$L_{\alpha\beta}^{{{\rm cl}}}$. 

For a systematic analysis of the general case, where \eqref{GKCCommH}
does not hold, we divide the perturbations
\begin{equation}\label{GKCDecompG}
g_{wj}(t)\equiv g_{wj}^{{{\rm cl}}}(t) + g_{wj}^{{{\rm qu}}}(t)
\end{equation}
into a classical part $g_{wj}^{{{\rm cl}}}(t)$ satisfying 
\eqref{GKCCommH} and a coherent part $g_{wj}^{{{\rm qu}}}(t)$, which
is purely non-diagonal in the unperturbed energy-eigenstates. 
By inserting this decomposition into \eqref{GKC} and exploiting the
properties  of the super-operators $\D^{0\dagger}_\nu$ arising from
\eqref{ApxGKCPropUnpert1}, we find
\begin{align}
L_{wj,wk}&= L_{wj,wk}^{{{\rm cl}}}+ L_{wj,wk}^{{{\rm qu}}},
& L_{wj,q\nu}&= L_{wj,q_\nu}^{{{\rm cl}}},\nonumber\\
L_{q\nu,wj}&= L_{q\nu,wj}^{{{\rm cl}}},
& L_{q\nu,q\mu}&= L_{q\nu,q\mu}^{{{\rm cl}}},
\label{GKCQCDecomp}
\end{align}
where the coefficients $L_{\alpha\beta}^{{{\rm cl}}}$ and 
$L_{\alpha\beta}^{{{\rm qu}}}$ are obtained by replacing $g_{wj}(t)$
with $g_{wj}^{{{\rm cl}}}(t)$ and $g_{wj}^{{{\rm qu}}}(t)$ in the
definitions \eqref{GKC}, respectively. 

This additive structure follows from a general argument, which we
provide in the second part of appendix~\ref{ApxQCS}.
It reveals two important features of the kinetic coefficients
\eqref{GKC}. 
First, the coefficients $L_{wj,wk}$ interrelating the perturbations
applied by different controllers decay into the quasi-classical part
$L^{{{\rm cl}}}_{wj,wk}$ and a quantum correction 
$L^{{{\rm qu}}}_{wj,wk}$. 
The latter contribution is thereby independent of the classical 
perturbations $g^{{{\rm cl}}}_{wj}(t)$ and accounts for coherences
between different eigenstates of $H^0$.
Second, the remaining coefficients are unaffected by the coherent
perturbations $g^{{{\rm qu}}}_{wj}(t)$ and thus, in general, 
constitute quasi-classical quantities.

\subsection{A Hierarchy of New Constraints}

The reciprocity relations \eqref{GKCRR} establish a link between the
kinetic coefficients describing a certain thermodynamic cycle and 
those corresponding to its time-reversed counterpart.
For an individual process determined by fixed driving protocols
$g_\alpha(t)$, these relations do, however, not provide any
constraints.
Still, the kinetic coefficients \eqref{GKC} are subject to a set of
bounds, which do not involve the reversed protocols and can be
conveniently summarized in form of the three conditions
\begin{equation}\label{GKCDefA}
\mathbb{A}\succeq 0, \quad
\mathbb{A}^{{{\rm cl}}} \succeq 0
\quad\text{and}\quad
\mathbb{A}-\mathbb{A}^{{{\rm cl}}}\succeq 0,
\end{equation}
where 
\begin{align}
\mathbb{A}&\equiv\frac{1}{2}
\left(\! \begin{array}{ccc}
2\mathbb{L}_{qq}^{{{\rm ins}}} & 2\mathbb{L}_{qw} & 2\mathbb{L}_{qq}\\
2\mathbb{L}_{qw}^t             
& \mathbb{L}_{ww}+\mathbb{L}_{ww}^t
& \mathbb{L}_{wq}+\mathbb{L}_{qw}^t\\
2\mathbb{L}_{qq}^t
& \mathbb{L}_{qw}+\mathbb{L}_{wq}^t
& \mathbb{L}_{qq}+\mathbb{L}_{qq}^t
\end{array}\!\right) \quad\text{and}\nonumber\\
\mathbb{A}^{{{\rm cl}}}  &\equiv
\bigl. \mathbb{A} \bigr|_{L_{wj,wk}\rightarrow 
L^{{{\rm cl}}}_{wj,wk}}.
\label{GKCDefA2}
\end{align}
Here, we used the block matrices $\mathbb{L}_{ab}$ introduced in 
\eqref{GTDDefBlockMat}, the diagonal matrix
\begin{equation}
\mathbb{L}^{{{\rm ins}}}_{qq}\equiv {{\rm diag}}\left(
\Li_{q1,q1},\dots, \Li_{qN_q,qN_q}\right)
\end{equation}
with entries defined in \eqref{GKC} and the quasi-classical kinetic 
coefficients $L_{wj,wk}^{{{\rm cl}}}$ introduced in 
\eqref{GKCQCDecomp}. 
Furthermore the notation $\bullet\succeq 0$ indicates that the 
matrices $\mathbb{A}$, $\mathbb{A}^{{{\rm cl}}}$ and $\mathbb{A}-
\mathbb{A}^{{{\rm cl}}}$ are positive semidefinite. 
The proof of this property, which we give in appendix \ref{ApxNC},
does not involve the $T$-symmetry relation \eqref{MDDetailedBalance}
but rather relies only on the condition \eqref{MDAdiabaticInv},
the detailed balance relation \eqref{MDDetailedBalanceLoc} and 
the corresponding structure \eqref{ApxGKCPropUnpert1} of the 
Lindblad-generator. 
We note that, in the classical realm, where $\mathbb{A}^{{{\rm
cl}}}=\mathbb{A}$, \eqref{GKCQCDecomp} reduces to the single condition
$\mathbb{A}\succeq 0$. 

The second law stipulates that the matrix $\mathbb{L}^{{{\rm s}}}$ 
defined in \eqref{GTDDefSymOM} must be positive semidefinite. 
Since $\mathbb{L}^{{{\rm s}}}$ is a principal submatrix of 
$\mathbb{A}$, this constraint is included in the first of the 
conditions \eqref{GKCDefA}, which thus explicitly confirms that our
formalism is thermodynamically consistent. 
Moreover, \eqref{GKCDefA} implies a whole hierarchy of constraints
on the generalized kinetic coefficients beyond the second law
\eqref{GTDDefSymOM}.
These bounds can be derived by taking successively larger principal
submatrices of $\mathbb{A}$, $\mathbb{A}^{{{\rm cl}}}$ or $\mathbb{A}
-\mathbb{A}^{{{\rm cl}}}$, which are not completely contained in 
$\mathbb{L}^{{{\rm s}}}$, and demanding their determinant to be 
non-negative. 
For example, by considering the principal submatrix 
\begin{equation}\label{GKCA2Constr}
\mathbb{A}_2^{{{\rm cl}}}\equiv\left(\!\begin{array}{cc}
2L_{wj,wj}^{{{\rm cl}}} & L_{wj,q\nu}+ L_{q\nu,wj}\\
L_{wj,q\nu}+ L_{q\nu,wj}& 2L_{q\nu,q\nu}
\end{array}\!\right)
\end{equation}
of $\mathbb{A}^{{{\rm cl}}}$ we find 
\begin{equation}\label{GKCA2CohConstr}
L_{wj,wj}^{{{\rm cl}}}
L_{q\nu,q\nu}-(L_{wj,q\nu}+L_{q\nu,wj})^2/4\geq 0. 
\end{equation}
Analogously, the principal submatrix 
\begin{equation}
\mathbb{A}_3^{{{\rm cl}}}\equiv \frac{1}{2} \left(\!\begin{array}{ccc}
2\Li_{q\nu,q\nu} & 2L_{q\nu,wj} & 2L_{q\nu,q\nu}\\
2L_{q\nu,wj}     & 2L_{wj,wj}^{{{\rm cl}}}   & 
L_{wj,q\nu} + L_{q\nu,wj}\\
2L_{q\nu,q\nu}   & L_{q\nu,wj}+L_{wj,q\nu} & 2L_{q\nu,q\nu}
\end{array}\!\right)
\end{equation}
yields the particularly important relation
\begin{equation}\label{GKCAConstr}
\frac{L_{q\nu,q\nu}}{\Li_{q\nu,q\nu}}\leq 
\frac{L_{wj,wj}^{{{\rm cl}}}
L_{q\nu,q\nu}-(L_{wj,q\nu}+L_{q\nu,wj})^2/4}{
L_{wj,wj}^{{{\rm cl}}}L_{q\nu,q\nu}-L_{wj,q\nu}L_{q\nu,wj}}.
\end{equation}
The classical version of this constraint has been previously used
to derive a universal bound on the power output of thermoelectric
\cite{Brandner2015} and cyclic Brownian \cite{Brandner2015f} heat
engines. 
As we will show in the next section, \eqref{GKCAConstr} implies that
cyclic quantum engines are subject to an even stronger bound. 

\section{Quantum Heat Engines}\label{SecQTM}
We will now show how the framework developed so far can be used to
describe the cyclic conversion of heat into work through quantum
devices. 
To this end, we focus on systems that are driven by a single external
controller with corresponding affinity $\F_w$ and one thermal force 
$\F_q$ such that two fluxes $J_w$ and $J_q$ emerge. 
For convenience, we omit the additional indices counting controllers
and reservoirs throughout this section. 
We note that this general setup covers not only heat engines but also
other types of thermal machines. 
An analysis of cyclic quantum refrigerators, for example, can be found
in appendix~\ref{ApxR}.

\subsection{Implementation}
A proper heat engine is obtained under the condition $J_w<0$, i.e.,
the external controller, on average, extracts the positive power
\begin{equation}\label{QTMPow}
P\equiv -\frac{1}{\T}\tint\tr{\dot{H}(t)\rc(t)} = - \Tc\F_w J_w
\end{equation}
per operation cycle while the system absorbs the heat flux $J_q>0$. 
The efficiency of this process can be consistently defined as
\cite{Brandner2015f}
\begin{equation}\label{QTMEff}
\eta\equiv P/J_q\leq\hc\equiv 1- \Th/\Tc,
\end{equation}
where the Carnot bound $\hc$ follows from the second law
$\dot{S}\geq 0$ and the bilinear form \eqref{GTDBilinEntropy} of the
entropy production. 
This figure generalizes the conventional thermodynamic efficiency
\cite{Callen1985}, which is recovered if the system is coupled to 
two reservoirs with respectively constant temperatures $\Tc$ and
$\Th$, either alternately or simultaneously. 
Both of these scenarios, for which $J_q$ becomes the average heat
uptake from the hot reservoir, are included in our formalism as 
special cases. 
The first one is realized by the protocol
\begin{equation}
\gamma_q(t)\equiv\begin{cases} 1 &\text{for}\quad 0\leq t< \T_1\\
                               0 &\text{for}\quad \T_1\leq t < \T
\end{cases}
\end{equation}
with $0<\T_1<\T$, the second one by setting $\gamma_q(t)= 1$. 

\subsection{Bounds on Efficiency and Power}
Optimizing the performance of a heat engine generally constitutes a
highly nontrivial task, which is crucially determined by the type of 
admissible control operations \cite{Seifert2012}. 
Following the standard approach, here we consider the thermal gradient
$\F_q$ and the temperature protocol $\gamma_q(t)$ as prespecified
\cite{Schmiedl2008,Holubec2014,Brandner2015f,Dechant2015,Bauer2016,
Dechant2016}. 
The external controller is allowed to adjust the strength of the 
energy modulation $\F_w$ and to select $g_w(t)$ from the space
of permissible driving protocols, which is typically restricted by 
natural limitations such as inaccessible degrees of freedom
\cite{Bauer2016}. 
Furthermore, we focus our analysis on the linear response regime, 
where general results are available due to fluxes and affinities
obeying the simple relations
\begin{equation}\label{QTMLinFlux}
J_w= L_{ww}\F_w + L_{wq}\F_q \quad\text{and}\quad
J_q= L_{qw}\F_w + L_{qq}\F_q.
\end{equation}

Rather than working directly with the kinetic coefficients showing 
up in \eqref{QTMLinFlux}, it is instructive to introduce the
dimensionless quantities
\begin{align}
x\equiv\frac{L_{wq}}{L_{qw}}, \quad
y\equiv\frac{L_{wq}L_{qw}}{L_{ww}L_{qq}-L_{wq}L_{qw}}, \quad
z\equiv\frac{L^{{{\rm qu}}}_{ww}L_{qq}}{L_{wq}^2}, 
\label{QTMDimlessPara}
\end{align}
which admit the following physical interpretation. 
First, we observe that, if the perturbations are invariant under full
time-reversal, i.e., if 
\begin{equation}\label{QTMSymProt}
g_{w}(t) = \TR g_{w}(-t)\quad\text{and}\quad
g_{q}(t) = \TR g_{q}(-t),
\end{equation}
the reciprocity relations \eqref{GKCRR} imply $L_{wq}=L_{qw}$ and 
thus $x=1$.
Thus, $x$ provides a measure for the degree, to which time-reversal
symmetry is broken by the external driving.
Second, $y$ constitutes a generalized figure of merit
accounting for dissipative heat losses. 
As a consequence of the second law, it is subject to the bound
\begin{equation}
h\leq y \leq 0\quad\text{for}\;\; x<0, \quad\text{and}\quad
0\leq y\leq h \quad\text{for}\;\; x\geq 0
\end{equation}
with $h\equiv 4x/(x-1)^2$ \cite{Benenti2011,Brandner2015f}. 
Third, the parameter $z$ quantifies the amount of coherence between
unperturbed energy eigenstates that is induced by the external 
controller. 
If $g_w(t)$ commutes with $H^0$, i.e., if the system behaves 
quasi-classically, the quantum correction $L^{{{\rm qu}}}_{ww}$
vanishes leading to $z=0$. 
Since $L_{qq},L_{ww}^{{{\rm qu}}}\geq 0$ by virtue of \eqref{GKCDefA}, 
for any proper heat engine, $z$ is strictly positive if the driving 
protocol is non-classical. 

We will now show that the presence of coherence profoundly impacts the
performance of quantum heat engines. 
In order to obtain a first benchmark parameter, we insert
\eqref{QTMLinFlux} into the definition \eqref{QTMEff} and take the
maximum with respect to $\F_w$. 
This procedure yields the maximum efficiency 
\begin{equation}\label{QTMMaxEff}
\eta_{{{\rm max}}}= \eta_{{{\rm C}}}
 x\frac{\sqrt{1+y}-1}{\sqrt{1+y}+1},
\end{equation}
which becomes equal to the Carnot value $\eta_{{{\rm C}}}=\Tc\F_q
+\mathcal{O}(\Delta T^2)$ in the reversible limit $y\rightarrow h$. 
However, the constraint \eqref{GKCA2CohConstr} stipulates 
\begin{equation}\label{QTMyCohCond}
h_z\leq y \leq 0 \;\;\text{for}\;\; x<0 \quad\text{and}\quad
0\leq y \leq h_z \;\;\text{for}\;\; x\geq 0
\end{equation}
with $h_z\equiv 4x/((x-1)^2 +4x^2 z)$ thus giving rise to the stronger
bound 
\begin{equation}\label{QTMMaxEffBnd}
\eta_{{{\rm max}}}\leq \eta_{{{\rm C}}} x 
\frac{\sqrt{1+h_z}-1}{\sqrt{1+h_z}+1}\leq 
\frac{\eta_{{{\rm C}}}}{1+4z},
\end{equation}
where the second inequality can be saturated only asymptotically for
$x\rightarrow\pm\infty$. 
This bound, which constitutes one of our main results, shows that
Carnot efficiency is intrinsically out of reach for any cyclic quantum
engine operated with a non-classical driving protocol in the linear 
response regime.  

As a second indicator of performance, we consider the maximum power
output 
\begin{equation}\label{QTMMaxPow}
P_{{{\rm max}}}=\frac{\Tc\F_q^2 L_{qq}}{4} \frac{xy}{1+y},
\end{equation}
which is found by optimizing \eqref{QTMPow} with respect to $\F_w$
using \eqref{QTMLinFlux}. 
This figure can be bounded by invoking the constraint
\eqref{GKCAConstr}, which, in terms of the parameters
\eqref{QTMDimlessPara}, reads 
\begin{equation}\label{QTMLqqConstrCoh}
L_{qq}\leq L_{qq}^{{{\rm ins}}}\frac{1-y/h_z}{1-xyz}.
\end{equation}
Replacing $L_{qq}$ in \eqref{QTMMaxPow} with this upper limit and 
maximizing the result with respect to $x$ and $y$ while taking into
account the condition \eqref{QTMyCohCond} yields 
\begin{equation}\label{QTMPowConstrCoh}
P_{{{\rm max}}}\leq \frac{\Tc\F_q^2L_{qq}^{{{\rm ins}}}}{4}
\frac{1}{1+z}.
\end{equation}
Hence, as a further main result, the power output is subject to an
increasingly sharper bound as the coherence parameter $z$ deviates 
from its quasi-classical value $0$. 
In the deep-quantum limit $z\rightarrow\infty$, which is realized if
the classical part $g_w^{{{\rm cl}}}(t)$ of the energy modulation
vanishes, both power and efficiency must decay to zero. 
These results hold under linear response conditions, however for any
temperature profile $\gamma_q(t)$ and any non-zero coherent driving 
protocol $g^{{{\rm qu}}}(t)$.

Finally, as an aside, we note that, even in the quasi-classical regime
the constraint \eqref{QTMLqqConstrCoh} rules out the option of Carnot 
efficiency at finite power, which, at least in principle, exists in 
systems with broken time-reversal symmetry  \cite{Benenti2011,
Brandner2013,Balachandran2013,Brandner2013a,Stark}. 
Specifically, for $z=0$, \eqref{QTMLqqConstrCoh} implies the relation
\cite{Brandner2015,Brandner2015f}
\begin{equation}\label{QTMGenConstr}
P\leq \Tc\F_q^2L_{qq}^{{{\rm ins}}} \begin{cases}
\frac{\eta}{\eta_{{{\rm C}}}}
\left(1-\frac{\eta}{\eta_{{{\rm C}}}}\right)
&\text{for}\quad |x|\geq 1\\[6pt]
\frac{\eta}{\eta_{{{\rm C}}}}\left(1-\frac{\eta}{\eta_{{{\rm C}}}x^2}
\right)
&\text{for}\quad |x|<1
\end{cases},
\end{equation}
which constrains the power output at any given efficiency $\eta$.
We leave the question how this detailed bound is altered when
coherence effects are taken explicitly into account as an interesting
subject for future research.

\section{Example}\label{SecEx}
\subsection{System and Kinetic Coefficients}
\begin{figure}
\epsfig{file=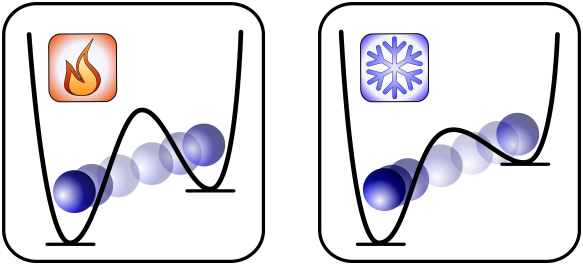,scale=0.068}
\caption{Two snapshots of the operation cycle of a two-level quantum
heat engine. 
A single particle is confined in a double well potential and
coupled to a thermal reservoir, whose temperature oscillates between
$\Th$ (left panel) and $\Tc<\Th$ (right panel). 
In a coarse-grained picture, this setup can be described as a
two-level system, where the particle is localized either in the left
or in the right well. 
Work is extracted from the system by varying a certain external 
control parameter, which affects both, the energetic difference 
between the two minima of the potential and the height of the barrier
separating them. 
This control operation, which corresponds to the non-classical driving
protocol \eqref{ExDrivProt}, inevitably allows the particle to tunnel
between the two wells. 
Consequently, it will typically be found in a coherent super-position
of the unperturbed energy-eigenstates during the thermodynamic cycle.
\label{Fig_TLS}}
\end{figure}

As an illustrative example for our general theory, we consider the 
setup sketched in Fig.~\ref{Fig_TLS}. 
A two-level system with free Hamiltonian 
\begin{equation}\label{ExUnpertH}
H^0 = \frac{\hbar\omega}{2}\sigma_z
\end{equation}
is embedded in a thermal environment, which is taken into account via
the unperturbed dissipation super-operator 
\begin{multline}
\D^0\bullet \equiv
\frac{\Gamma}{2}\left([\sigma_-\bullet,\sigma_+]
+[\sigma_-,\bullet\sigma_+]\right)\\
+\frac{\Gamma e^{-2\kappa}}{2}
\left([\sigma_+\bullet,\sigma_-]+[\sigma_+,\bullet\sigma_-]\right)
\end{multline}
with the dimensionless parameter 
\begin{equation}
\kappa \equiv \hbar\omega/(2\kb\Tc)
\end{equation}
corresponding to the rescaled level splitting. 
This system is driven by the temperature profile 
\begin{equation}
T(t)\equiv \frac{\Th\Tc}{\Th + (\Tc-\Th)\gamma_q(t)}.
\end{equation}
Simultaneously, work can be extracted through the energy modulation 
\begin{equation}\label{ExDrivProt}
\Delta H g_w(t) \equiv \Delta H \gamma_w(t)
\left(\cos\theta\;\sigma_z + \sin\theta\;\sigma_x\right),
\end{equation}
where $\gamma_w(t)$ and $\gamma_q(t)$ are $\T$-periodic functions of
time. 
Furthermore, $\sigma_x,\sigma_y,\sigma_z$ denote the usual Pauli
matrices and $\sigma_\pm\equiv (\sigma_x\pm i\sigma_y)/2$.
The parameter $0\leq\theta\leq\pi$ quantifies the relative degree, to
which the external controller induces shifting of the free energy 
levels and coherent mixing between them. 

The kinetic coefficients describing the thermodynamics of this system
in the linear response regime can be obtained from the relation 
\eqref{GKCSimplified}. 
To this end, we first evaluate the deviations from equilibrium  
\begin{align}
\delta g_w(t)&= \gamma_w(t) \left(\cos\theta\;\sigma_z 
+\tanh\kappa\cos\theta\;\mathbbm{1}+\sin\theta\;\sigma_x\right)
\nonumber,\\
\delta g_q(t)&= -\kb\Tc \kappa \gamma_q(t)\left(\sigma_z 
+\tanh\kappa\;\mathbbm{1}\right)
\end{align}
according to the definition \eqref{GKCEqFluct}.  
Inserting these expressions into \eqref{GKCSimplified}, after some 
straightforward algebra, yields
\begin{align}
L^{{{\rm cl}}}_{ab} &= -\frac{\xi^{{{\rm cl}}}_a\xi^{{{\rm cl}}}_b}
{\kb\T}\tint\Bigl(\dot{\gamma}_a(t)\gamma_b(t)\nonumber\\
&\hspace*{3.3cm}
-\tauint\dot{\gamma}_a(t)\dot{\gamma}_b(t-\tau)e^{-\hat{\Gamma}\tau}
\Bigr),\nonumber\\
L_{ww}^{{{\rm qu}}} &= \frac{(\xi_w^{{{\rm qu}}})^2}{\kb\T}
\ttauint \dot{\gamma}_w(t)\dot{\gamma}_w(t-\tau)e^{-\hat{\Gamma}\tau/2}
\cos[\omega\tau],
\label{ExGKC}
\end{align}
where $a,b=w,q$ and the abbreviations
\begin{align}
\xi_w^{{{\rm cl}}} &\equiv \cos\theta/\cosh\kappa, & 
\xi_w^{{{\rm qu}}} &\equiv 2\sqrt{\kappa\tanh\kappa}\sin\theta,\nonumber\\
\xi_q^{{{\rm cl}}} &\equiv -\kb\Tc\kappa/\cosh\kappa, &
\hat{\Gamma} &\equiv \Gamma(1+e^{-2\kappa})
\end{align}
were introduced for convenience. 
We note that, obviously, these coefficients fulfill the symmetry 
relation \eqref{GKCRRFact} due to the driving protocol
\eqref{ExDrivProt} satisfying the factorization condition
\eqref{GKCFactCond}. 
Finally, for later purposes, we evaluate the instantaneous coefficient
\begin{equation}\label{ExLqqIns}
L^{{{\rm ins}}}_{qq} = \frac{(\xi_q^{{{\rm cl}}})^2\hat{\Gamma}}{\kb\T}
\tint \gamma_q^2(t),
\end{equation}
which is defined in \eqref{GKC} and enters the constraint 
\eqref{QTMPowConstrCoh}.

\subsection{Power and Efficiency}\label{SubSecExHm}

We will now explore the performance of the toy model of 
Fig.~\ref{Fig_TLS} as a quantum heat engine. 
In order to keep our analysis as simple and transparent as possible,
we assume harmonic protocols
\begin{align}
\gamma_w(t) &= \sin\left[2\pi t/\T + \phi\right]
\quad\text{and}\nonumber\\
\gamma_q(t) &= \left(1+ \sin\left[2\pi t/\T\right]\right)/2,
\label{ExHmDrivProt}
\end{align}
where the phase shift $\phi$ can be adjusted to optimize the device
for a given purpose.
The kinetic coefficients \eqref{ExGKC} and \eqref{ExLqqIns} then
become 
\begin{align}
L_{ww} &= L_{ww}^{{{\rm cl}}} + L_{ww}^{{{\rm qu}}} 
=\frac{(\xi^{{{\rm cl}}}_w)^2}{\kb\T}\frac{\pi\alpha}{1+\alpha^2}
\nonumber\\
       &\hspace*{1.035cm}+ \frac{(\xi^{{{\rm qu}}}_w)^2}{\kb\T}
          \frac{2\pi\alpha\nu^2(4\nu^2+\alpha^2(\nu^2+1))}
               {16\nu^4+8\alpha^2\nu^2(\nu^2-4)+\alpha^4(\nu^2+4)^2},
\nonumber\\
L_{wq} &= L_{wq}^{{{\rm cl}}} =  
\frac{\xi^{{{\rm cl}}}_w\xi^{{{\rm cl}}}_q}{\kb\T}
\frac{\pi\alpha}{1+\alpha^2}\frac{\cos\phi+\alpha\sin\phi}{2},
\nonumber\\
L_{qw} &= L_{qw}^{{{\rm cl}}} = 
\frac{\xi^{{{\rm cl}}}_w\xi^{{{\rm cl}}}_q}{\kb\T}
\frac{\pi\alpha}{1+\alpha^2}\frac{\cos\phi-\alpha\sin\phi}{2},
\nonumber\\
L_{qq} &= L_{qq}^{{{\rm cl}}} =
\frac{(\xi^{{{\rm cl}}}_q)^2}{4\kb\T}\frac{\pi\alpha}{1+\alpha^2}
\label{ExHmGKC}
\end{align}
and
\begin{equation}\label{ExHmLqqins}
L_{qq}^{{{\rm ins}}} = \frac{(\xi_q^{{{\rm cl}}})^2}{\kb\T}
\frac{3\pi\alpha}{4},
\end{equation}
respectively, with 
\begin{equation}
\alpha\equiv\hat{\Gamma}\T/2\pi \quad\text{and}\quad
\nu\equiv \hat{\Gamma}/\omega
\end{equation}
being dimensionless constants. 

\begin{figure}
\centering 
\epsfig{file=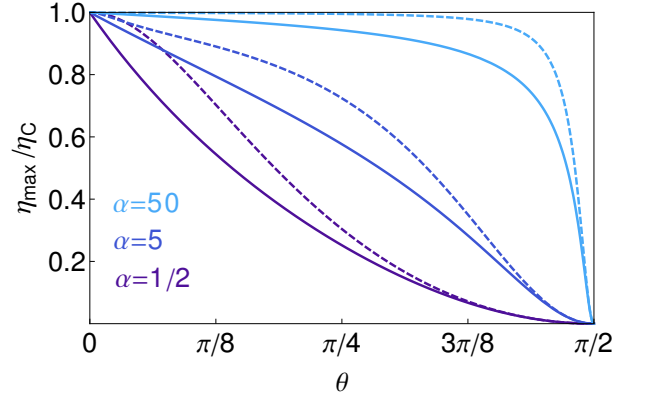,scale=0.65}
\caption{Maximum efficiency of a two-level quantum heat engine. 
The solid lines show the explicit result \eqref{ExHmMaxEff} in units
of the Carnot efficiency $\eta_{{{\rm C}}}$ as function of the
coherence parameter $\theta$ for different values of the damping
parameter $\alpha$ . 
The dashed lines indicate the corresponding bound \eqref{QTMMaxEffBnd}
evaluated with the protocols \eqref{ExHmDrivProt} and the optimal 
phase shift \eqref{ExHmMaxEffProt}. 
The remaining parameters have been chosen as $\kappa=1/2$ and 
$\nu=10$.
For clarity, the legend in the lower left corner follows the order
of the plotted curves from top to bottom. 
\label{FigExHmEff}}
\end{figure}
Within these specifications, the maximal efficiency is found by
inserting \eqref{ExHmGKC} into \eqref{QTMDimlessPara} and
\eqref{QTMMaxEff} and taking the maximum with respect to $\phi$. 
This procedure yields
\begin{equation}\label{ExHmMaxEff}
\eta_{{{\rm max}}}=\eta_{{{\rm C}}}
\frac{\alpha\psi_1-\psi_2}{\alpha\psi_1+\psi_2}
\end{equation}
and the corresponding optimal phase shift 
\begin{equation}\label{ExHmMaxEffProt}
\phi_\eta = \arccos [(\psi_1^2-\psi_2^2)/(\psi_1^2+\psi_2^2)]/2,
\end{equation}
where
\begin{align}
\psi_1 &\equiv\sqrt{2\kb\T L^{{{\rm qu}}}_{ww}
+2\pi\alpha(\xi^{{{\rm cl}}}_w)^2}\quad\text{and}\nonumber\\
\psi_2 &\equiv\alpha\sqrt{2\kb\T L^{{{\rm qu}}}_{ww}}.
\end{align}
In the quasi-classical limit, where $L^{{{\rm qu}}}_{ww}$ and 
thus $\psi_2=0$, these expressions reduce to 
\begin{equation}
\left.\eta_{{{\rm max}}}\right|_{\theta=0}=\eta_{{{\rm C}}} 
\quad\text{and}\quad
\left.\phi_\eta\right|_{\theta=0}= 0. 
\end{equation}
Hence, the engine can indeed reach Carnot efficiency if the protocols
$\gamma_w(t)$ and $\gamma_q(t)$ are in phase with each other. 
As Fig.~\ref{FigExHmEff} shows, the maximum efficiency falls monotonically from $\eta_{{{\rm C}}}$ to $0$ as $\theta$ varies from
$0$ to $\pi/2$.
Moreover, the decay proceeds increasingly faster the smaller the
damping parameter $\alpha$ is chosen. 
This observation can be understood intuitively, since, for large 
$\alpha$, the thermodynamic cycle evolves close to the adiabatic
limit, where it becomes reversible. 
As a reference point, the bound \eqref{QTMMaxEffBnd} has been included
in Fig.~\ref{FigExHmEff}. 
It shows the same qualitative dependence on $\theta$ and $\alpha$ as 
the maximum efficiency, for which it provides a fairly good estimate,
especially as $\theta$ comes close to $\pi/2$. 

\begin{figure}
\centering
\epsfig{file=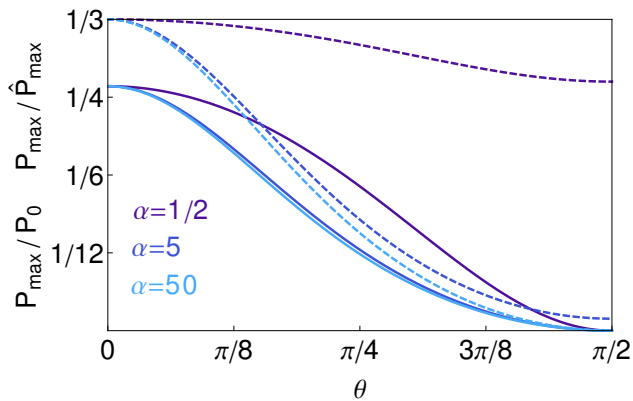,scale=0.65}
\caption{Dependence of the maximum power of a two-level quantum engine
on the coherence parameter $\theta$ for $\kappa=1/2$, $\nu=10$ and 
three different values of $\alpha$. 
The solid lines correspond to the optimized output \eqref{ExHmMaxPow}
in units of $P_0\equiv \Tc\F_q^2(\xi^{{{\rm cl}}})^2\alpha/(12\kb\T)$
\cite{Note2}.
The dashed lines show the maximum power as a fraction of its upper 
bound \eqref{ExHmMaxPowBnd}.
In the limit $\theta\rightarrow\pi/2$, both quantities,
$P_{{{\rm max}}}$ and the bound $\hat{P}_{{{\rm max}}}$, vanish, while
their ratio approaches a finite value. 
The legend in the lower left corner has been sorted according to the 
order of the plotted curves from top to bottom. 
This correspondence applies to dashed and solid lines, respectively.
\label{FigExHmPow}}
\end{figure}

We now turn to maximum power as a second important benchmark 
parameter. 
Combining \eqref{ExHmGKC}, \eqref{QTMDimlessPara} and
\eqref{QTMMaxPow}, after maximization with respect to $\phi$, yields
the explicit expression
\begin{equation}\label{ExHmMaxPow}
P_{{{\rm max}}}=\frac{\Tc\F_q^2}{4}\frac{\pi^2\alpha^2 
(\xi^{{{\rm cl}}}_q\xi^{{{\rm cl}}}_w)^2}{2\kb\T(\psi_1^2+\psi_2^2)},
\end{equation} 
where the optimal phase shift 
\begin{equation}\label{ExHmMaxPowPS}
\phi_{{{\rm P}}}= \arctan\alpha
\end{equation}
is independent of $\theta$. 
This result can be quantitatively assessed by comparing it with the
bound 
\begin{equation}\label{ExHmMaxPowBnd}
\hat{P}_{{{\rm max}}}=\frac{\Tc\F_q^2}{4}\frac{3\pi^2\alpha^2 
(\xi^{{{\rm cl}}}_q\xi^{{{\rm cl}}}_w)^2}{2\kb\T\psi_1^2},
\end{equation}
which follows from \eqref{QTMPowConstrCoh} after inserting
\eqref{ExHmLqqins} and evaluating the parameter $z$ using the 
protocols \eqref{ExHmDrivProt} with $\phi=\phi_{{{\rm P}}}$. 

In Fig.~\ref{FigExHmPow}, both, the optimal power \eqref{ExHmMaxPow} 
and the ratio 
\begin{equation}\label{ExHmPowRatio}
\frac{P_{{{\rm max}}}}{\hat{P}_{{{\rm max}}}}
=\frac{\psi_1^2}{3(\psi_1^2+\psi_2^2)} 
\end{equation}
are plotted. 
Two central features of these quantities are can be observed. 
First $P_{{{\rm max}}}$ reaches its maximum as a function of
$\theta$ in the quasi-classical case $\theta=0$ and then decays
monotonically to zero as $\theta$ approaches $\pi/2$
\footnote{Note that the standard power $P_0$ contains a factor 
$\alpha$ for the following reason. 
The bare maximum power \eqref{ExHmMaxPow} grows linearly in
$\alpha$ and can, seemingly, become arbitrary large. 
However, through the optimization procedure leading to
\eqref{QTMMaxPow} the affinity $\F_w$ has been fixed as
$\F_w=-\F_q L_{wq}/(2L_{ww})$. 
It is straightforward to check that, for $\alpha\gg 1$, the ratio of
kinetic coefficients showing up here becomes proportional to $\alpha$
if the phase shift \eqref{ExHmMaxPowPS} is chosen. 
Thus, in order to stay within the linear response regime, 
$\F_q$ must be assumed inversely proportional to $\alpha$ such that
the power output is effectively bounded.}.
This behavior is in line with our general insight that coherence 
effects are detrimental to the performance of quantum heat engines. 
Second, in contrast to maximum efficiency, the maximum power comes 
not even close to the upper limit following from our new constraint 
\eqref{GKCDefA}. 
Specifically, the degree of saturation \eqref{ExHmPowRatio} is equal to 
$1/3$ for $\theta=0$ and then decreases even further towards
$\theta=\pi/2$. 
Still, the bound \eqref{QTMPowConstrCoh} might be attainable by more
complex devices than the one considered here. 
Whether or not such models exist remains an open question at this 
point.

\begin{widetext}
\begin{center}
\begin{figure}
\epsfig{file=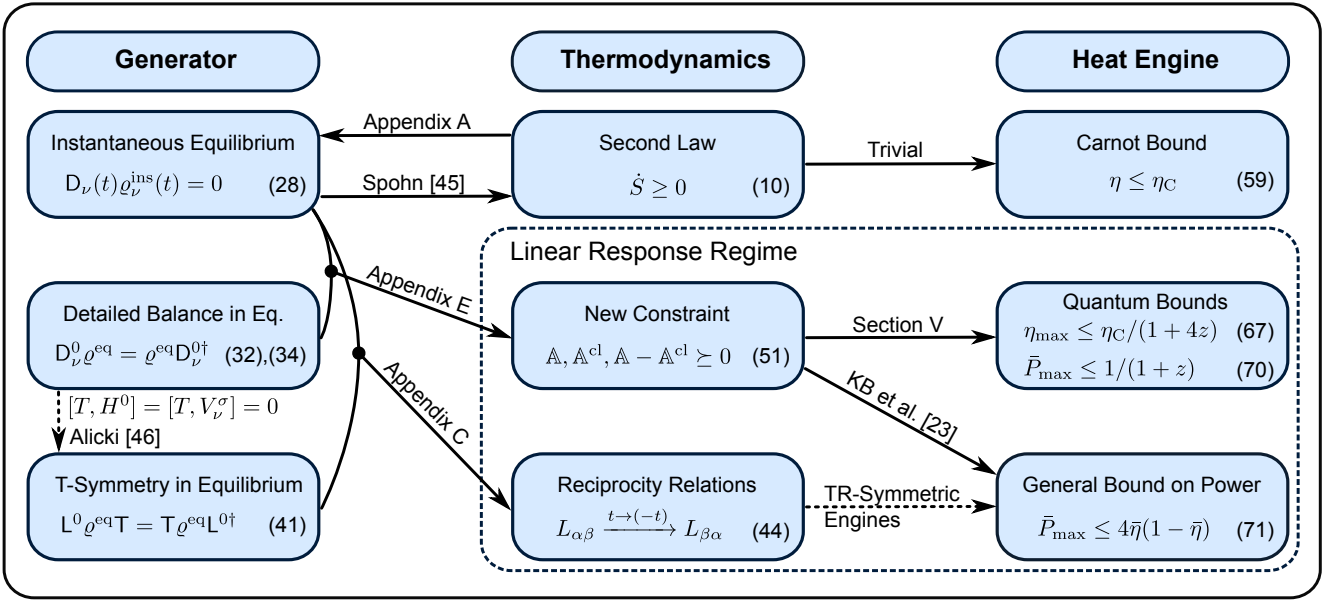,scale=2.0}
\caption{Flow chart visualizing the interdependence between 
properties of the Lindblad-generator (left column), relations between 
thermodynamic quantities (central column) and bounds on the 
performance figures of quantum heat engines (right column). 
Solid arrows denote unrestricted implications, while dashed arrows
require the additional condition attached to them. 
In the last column, we used the abbreviations $\bar{\eta}\equiv
\eta/\eta_{{{\rm C}}}$ and $\bar{P}_{{{\rm max}}}\equiv 
P_{{{\rm max}}}/P_0$, where $P_0=\Tc\F_q^2 L_{qq}/4$ for the dashed
arrow in the bottom line and otherwise 
$P_0=\Tc\F_q^2 L_{qq}^{{{\rm ins}}}/4$. 
An engine is considered to be time-reversal (TR) symmetric if the
corresponding driving protocols fulfill the condition 
\eqref{QTMSymProt}.
\label{FigFlowChart}}
\end{figure}
\end{center}
\end{widetext}

\section{Concluding Perspectives}\label{SecCon}

In this paper, we have developed a universal framework for the
description of quantum thermodynamic cycles, which allows the
consistent definition of kinetic coefficients relating fluxes and
affinities for small driving amplitudes. 
Focusing on Markovian dynamics, we have proven that these quantities
fulfill generalized reciprocity relations and, moreover, are subject
to a set of additional constraints.
These results were derived from the characteristics of the 
Lindblad-generator as summarized in Fig.~\ref{FigFlowChart}. 
To this end, we have invoked two fundamental physical principles. 
First, in order to ensure consistency with the second law, each 
dissipation super-operator must annihilate the instantaneous 
Gibbs-Boltzmann distribution at the respectively corresponding
temperature. 
Second, we have demanded the dissipative parts of the unperturbed
generator to fulfill a detailed balance relation implying zero 
probability flux between any pair or energy eigenstates in 
equilibrium. 
For the reciprocity relations, the even stronger $T$-symmetry 
condition is necessary. 
Both, detailed balance and $T$-symmetry are quite natural and 
broadly accepted conditions, which ultimately rely on the 
reversibility of microscopic dynamics.  
It should, however, be noted that, at least from a 
phenomenological point of view, they constitute stronger
requirements than the bare second law, which stipulates only the 
first of the above mentioned properties of the Lindblad generator.

As a key application, our theory allows to obtain bounds on the 
maximum efficiency and power of quantum heat engines, which reveal
that coherence effects are generally detrimental to both of these
figures of merit.  
This insight has been illustrated quantitatively for a 
paradigmatic model consisting of a harmonically driven two-level 
system. 
In the quasi-classical limit, where our new constraints on the 
kinetic coefficients become weakest, we recover a general bound
on power, which is a quadratic function of efficiency. 
This relation, which has been derived before for classical 
stochastic \cite{Brandner2015f} and thermoelectric heat
engines \cite{Brandner2015}, in particular proves the 
nonexistence of reversibly operating quantum devices with finite power
output, at least within linear response. 
For classical systems, the analogous result was obtained also in 
\cite{Proesmans2015,Proesmans2015a} and, only recently, extended to
the more general nonlinear regime in \cite{Shiraishi2016}.
All of these approaches, however, rely on a Markovian dynamics, which
is further specified by a detailed balance condition. 
Since, as we argued before, this requirement is more restrictive when
demanding only the  non-negativity of entropy production, the 
incompatibility of Carnot efficiency and finite power can not be
attributed to the bare second law. 

Despite the fact that our discussion has mainly focused on quantum
heat engines, it is clear that our general framework covers also other
types of thermal machines like, for  example, quantum absorption
refrigerators \cite{Levy2012,Correa2014,Mitchison2015a}. 
It can be expected that the new constraints on the kinetic coefficients 
derived here allow to restrict also the figures of performance of such
devices.  
Working out these bounds explicitly is left as an interesting topic for 
future research at this point. 

Analyses of the linear response regime can provide profound insights
on the properties of non-equilibrium systems.
A complete understanding of their behavior, however, typically 
requires to take strong-driving effects into account. 
Quantum heat engines, for example, that are operated by purely 
non-classical protocols do not admit a proper linear response 
description, since their off-diagonal kinetic coefficients would 
inevitably vanish. 
A paradigmatic model belonging to this class is, for example, the
coherently driven three-level amplifier
\cite{Scovil1959,Geva1994,Geva1996}.
It thus emerges the question how our new constraint \eqref{GKCDefA}
and thus the bounds \eqref{QTMMaxEffBnd}, \eqref{QTMPowConstrCoh} and
\eqref{QTMGenConstr} can be extended to the nonlinear regime. 
Investigations towards this direction constitute an important topic, 
which can be expected to be challenging, since universal results
for systems arbitrary far from equilibrium are overall scarce. 
Indeed, the general framework of Sec.~\ref{SecFramework} is
not tied to the assumption of small driving amplitudes. 
However, accounting for strong  perturbations, might, for example,
require to specify the dynamical generator in a more restrictive way 
when it was done in Sec.~\ref{SecMarkovDyn} thus sacrificing
universality. 

In summary, our approach provides an important first step towards a
systematic theory of cyclic quantum thermodynamic processes. 
It should thus provide a fruitful basis for future investigations,
which could eventually lead to a complete understanding of the
fundamental principles governing the performance of quantum thermal
devices. 

\begin{acknowledgments}
K.B. was supported by the Academy of Finland Centre of Excellence 
program (project 284594).
\end{acknowledgments}

\appendix

\section{Thermodynamic Consistency of the Time-Dependent Lindblad Equation}\label{SecSecondLaw}

We consider the total rate of entropy production 
\eqref{GTDEntropyProd}, which can be rewritten as 
\begin{align}
\dot{S}[\varrho(t)] &= -\kb \tr{\dot{\varrho}(t)\ln\varrho(t)}
        -\sum_{\nu=1}^{N_q}\frac{\dot{Q}_\nu(t)}{T_\nu(t)}\nonumber\\
&= -\kb\sum_{\nu=1}^{N_q}\tr{\bigl(\D_\nu(t)\varrho(t)\bigr)
    \bigl(\ln\varrho(t)-\ln\varrho_\nu^{{{\rm ins}}}(t)\bigr)}
    \nonumber\\
&\equiv \sum_{\nu=1}^{N_q} \dot{S}_\nu[\varrho(t)].
\label{ApxTDCSDef}
\end{align}
As proven by Spohn \cite{Spohn1978}, the condition 
\eqref{MDAdiabaticInv} is sufficient for each of the contributions 
$S_\nu[\varrho(t)]$ to be non-negative for any $\varrho(t)$. 
Here, we show that \eqref{MDAdiabaticInv} is also necessary to this 
end. 

We proceed as follows. 
First, we define a one-parameter family of states 
$\varrho^\lambda_\nu$ such that $\varrho^{\lambda=0}_\nu=
\varrho^{{{\rm ins}}}_\nu$ and $\dot{S}_\nu[\varrho_\nu^\lambda]$
at least once continuously-differentiable at $\lambda=0$.  
Hence, we obviously have
\begin{equation}
\left. \dot{S}_\nu[\varrho_\nu^\lambda]\right|_{\lambda=0} = 0. 
\end{equation}
Note that, for convenience, we omit time-arguments from here 
onwards.
Second, we observe that, due to continuity, the family 
$\varrho_\nu^\lambda$ will always contain a state
$\varrho^{\lambda\ast}_\nu$ in the vicinity of $\lambda=0$ such that
$\dot{S}_\nu[\varrho^{\lambda\ast}_\nu]< 0$ unless 
\begin{equation}\label{ApxTDCDevS}
\left.\partial_\lambda
\dot{S}_\nu[\varrho_\nu^\lambda]\right|_{\lambda=0}=0.
\end{equation}
Third, we set
\begin{equation}\label{ApxTDCSpecFam}
\varrho_\nu^\lambda = \exp[-H/(\kb T_\nu)+\lambda G]/Z(\lambda),
\end{equation}
where $Z(\lambda)\equiv\tr{\exp[-H/(\kb T_\nu)+\lambda G]}$ and 
$G$ is an arbitrary Hermitian operator. 
Inserting \eqref{ApxTDCSpecFam} into \eqref{ApxTDCDevS} and 
using \eqref{ApxTDCSDef} and \eqref{MDDissipator} yields 
\begin{equation}
\tr{\bigl(\D_\nu\varrho^{{{\rm ins}}}_\nu\bigr)G}=0.
\end{equation}
Finally, this condition can only be satisfied for any Hermitian
$G$ if $\D_\nu\varrho^{{{\rm ins}}}_\nu=0$. 
Thus, we have shown that, if \eqref{MDAdiabaticInv} is not 
fulfilled, we can always construct a state 
$\varrho_\nu^{\lambda\ast}$ such that 
$\dot{S}_\nu[\varrho_\nu^{\lambda\ast}]$ becomes
negative, which completes the proof.

\section{Generalized Kinetic Coefficients}\label{ApxGKC}
\subsection{General Set-up}
We derive the expressions \eqref{GKC} for the generalized kinetic
coefficients within three steps. 
First, by linearizing the components of the generator 
\eqref{MDGenerator} with respect to $\Delta_j H$ and $\Delta_\nu T$,
we obtain 
\begin{align}
\H(t)& \equiv \H^0 + \sum_{j=1}^{N_w} \Delta_jH \H^{w j}(t)
+\mathcal{O}(\Delta^2),
\nonumber\\
\D_\nu(t) &\equiv \D^0_\nu 
                +\sum_{j=1}^{N_w} \Delta_j H \D^{wj}_\nu(t)
                +\Delta_\nu T \D^{q}_\nu(t)
                +\mathcal{O}(\Delta^2),
\label{ApxGKCExpGen}
\end{align}
where we assume that $\D_\nu(t)$ depends on $H(t)$ and $T_\nu(t)$ but
not on $T_\mu(t)$ if $\mu\neq\nu$. 
The quantities showing up in these expansions can be characterized as 
follows. 
A straightforward calculation shows that the structure 
\eqref{ApxGKCPropUnpert1} implies
\begin{equation}\label{ApxGKCCummutRel}
\D^0_\nu\lint R^\lambda\bullet R^{-\lambda}\req
= \lint R^\lambda\left(\D^{0\dagger}_\nu\bullet\right)
R^{-\lambda}\req,
\end{equation}
where
\begin{multline}\label{ApxGKCDBAdjDiss}
\D^{0\dagger}_\nu\bullet = \frac{1}{2}\sum_\sigma
\Gamma^\sigma_\nu\left(V_\nu^\sigma[\bullet,V^{\sigma\dagger}_\nu]
+ [V^\sigma_\nu,\bullet]V^{\sigma\dagger}_\nu\right)\\
+ \bar{\Gamma}^\sigma_\nu
\left(V_\nu^{\sigma\dagger}[\bullet,V^\sigma_\nu]
+ [V^{\sigma\dagger}_\nu,\bullet]V^\sigma_\nu\right).
\end{multline}
Furthermore, by expanding the relation \eqref{MDAdiabaticInv} to 
linear order in $\Delta_j H$ and $\Delta_\nu T$, we find 
\begin{align}
\D^{wj}_\nu(t)\req &= \frac{1}{\kb\Tc}\D^0_\nu\lint
R^{\lambda}g_{wj}(t)R^{-\lambda}\req,\nonumber\\
&= \frac{1}{\kb\Tc}\lint
R^{\lambda}\left(\D^{0\dagger}_\nu g_{wj}(t)\right)R^{-\lambda}
\req,\nonumber\\
\D^q_\nu\req &= \frac{1}{\kb(\Tc)^2}\D^0_\nu\lint 
R^\lambda g_{q\nu}(t)R^{-\lambda}\req\nonumber\\
&= \frac{1}{\kb(\Tc)^2}\lint R^\lambda\left(\D^{0\dagger}_\nu
g_{q\nu}(t)\right)R^{-\lambda}\req. 
\label{ApxGKCPropUnpert4}
\end{align}
Analogously, the trivial relation
\begin{equation}
\H(t)\exp[-H(t)/(\kb T_\nu(t))]=0
\end{equation}
yields
\begin{align}
\H^{wj}(t)\req & = \frac{1}{\kb\Tc}\H^0\lint R^{\lambda} g_{wj}(t)
R^{-\lambda}\req \nonumber\\
& = \frac{1}{\kb\Tc}\lint R^{\lambda} 
\left(\H^0 g_{wj}(t)\right)R^{-\lambda}\req.
\label{ApxGKCPropUnpert5}
\end{align}

As the second step of our derivation, we parametrize the density 
matrix $\rc(t)$ describing the limit-cycle of \eqref{MDLindblad} as
\begin{align}
\rc(t) & \equiv \frac{1}{Z^0}\exp\biggl[-\frac{H^0}{\kb\Tc}
+ \sum_{j=1}^{N_w} \frac{\Delta_j H}{\kb\Tc} G_{wj}(t)\nonumber\\
& \hspace*{2.7cm}
+ \sum_{\nu=1}^{N_q}\frac{\Delta_\nu T}{\kb(\Tc)^2} G_{q\nu}(t)
+\mathcal{O}\left(\Delta^2\right) \biggr]                     
\nonumber\\
& = \req +\sum_{j=1}^{N_w}\frac{\Delta_j H}{\kb\Tc}
    \lint R^\lambda G_{wj}(t) R^{-\lambda}\req \nonumber\\
& \hspace*{0.3cm}
+ \sum_{\nu=1}^{N_q}\frac{\Delta_\nu T}{\kb(\Tc)^2}\lint R^\lambda
G_{q\nu}(t) R^{-\lambda}\req +\mathcal{O}\left(\Delta^2\right).
\label{ApxGKCExpLimitCycle}
\end{align}
Inserting this expansion, \eqref{MDGenerator} and 
\eqref{ApxGKCExpGen} into \eqref{MDLindblad} and applying the 
relation \eqref{ApxGKCCummutRel} yields
\begin{align}
\partial_t G_{wj}(t) & = \K G_{wj}(t)+\K g_{wj}(t),\nonumber\\
\partial_t G_{q\nu}(t) &= \K G_{q\nu}(t)+
\D^{0\dagger}_\nu g_{q\nu}(t). 
\label{ApxGKCDiffEq}
\end{align}
By solving these differential equations with respect to the periodic 
boundary conditions $G_{wj}(t+\T)=G_{wj}(t)$ and $G_{q\nu}(t+\T)
=G_{q\nu}(t)$, we obtain
\begin{align}
G_{wj}(t) &=\tauint e^{\K\tau}\K g_{wj}(t-\tau),\nonumber\\
G_{q\nu}(t) &= \tauint e^{\K\tau}\D^{0\dagger}_\nu g_{q\nu}(t-\tau). 
\label{ApxGKCSolPertLimitCycle}
\end{align}
The integrals with infinite upper bound showing up in these 
expressions converge, since, due to the set of unperturbed
Lindblad-operators $\{V^\sigma_\nu\}$ being self-adjoint and
irreducible, the non-vanishing eigenvalues of $\K$ have negative 
real part \cite{Spohn1977}.
Moreover, $\mathbbm{1}$ is the unique right-eigenvector of $\K$ 
corresponding to the eigenvalue $0$. 
In \eqref{ApxGKCSolPertLimitCycle}, the super-operator $e^{\K\tau}$,
however, acts on operators, which, by construction, are linearly
independent of $\mathbbm{1}$, since $\D^{0\dagger}\mathbbm{1}=0$ and 
$\H^0\mathbbm{1}=0$. 
The same argument ensures that the general expressions \eqref{GKC} for
the kinetic coefficients are well-defined. 

For the third step, we recall the definitions \eqref{MDWorkFlux} and
\eqref{MDHeatFlux} of the generalized fluxes, 
\begin{align}
J_{wj}   &=-\frac{1}{\T}\tint \tr{g_{wj}(t)\L(t)\rc(t)}
       \quad\text{and}\label{ApxGKCWorkFlux}\\
J_{q\nu} &= \frac{1}{\T}\tint\gamma_{q\nu}(t)\tr{
H(t)\D^\dagger_\nu(t)\rc(t)}.
\label{ApxGKCHeatFlux}
\end{align}
Inserting \eqref{MDGenerator}, \eqref{ApxGKCExpGen},
\eqref{ApxGKCPropUnpert4}, \eqref{ApxGKCPropUnpert5} and 
\eqref{ApxGKCExpLimitCycle} into \eqref{ApxGKCWorkFlux},
neglecting all contributions of second order in $\Delta$ and applying
\eqref{ApxGKCCummutRel} leads to the generalized kinetic coefficients 
\begin{align}
L_{wj,wk}&= \frac{(-1)}{\kb\T}\tint\ev{g_{wj}(t)}{
              \K G_{wk}(t) + \K g_{wk}(t)},\nonumber\\
L_{wj,q\nu}&= \frac{(-1)}{\kb\T}\tint\ev{g_{wj}(t)}{
              \K G_{q\nu}(t) + \D^{0\dagger}_\nu g_{q\nu}(t)}.
\label{ApxGKCWorkCoeff}
\end{align}
Analogously, we obtain from \eqref{ApxGKCHeatFlux}
\begin{align}
L_{q\nu,wj}&= \frac{(-1)}{\kb\T}\tint\ev{g_{q\nu}(t)}{
\D^{0\dagger}_\nu G_{wj}(t) + \D^{0\dagger}_\nu g_{wj}(t)},\nonumber\\
L_{q\nu,q\mu} &= \frac{(-1)}{\kb\T}\tint\ev{g_{q\nu}(t)}{
\D^{0\dagger}_\nu G_{q\mu}(t) + \delta_{\nu\mu}\D^{0\dagger}_\nu
g_{q\nu}(t)}.
\label{ApxGKCHeatCoeff}
\end{align}
Finally, eliminating  $G_{wj}(t)$ and $G_{q\nu}(t)$ from 
\eqref{ApxGKCWorkCoeff} and \eqref{ApxGKCHeatCoeff} using 
\eqref{ApxGKCSolPertLimitCycle} gives the desired expressions 
\eqref{GKC}. 

\subsection{Simplified Set-up}
We consider the special case, where the system is attached only to a
single reservoir. 
In order to derive the simplified expressions \eqref{GKCSimplified}
for the generalized kinetic coefficients, we first note that, since
$\H^0 g_{q1}(t)=0$, we can replace $\D^{0\dagger}_1 g_{q1}(t)$ by 
$\K g_{q1}(t)$ in \eqref{GKC}.
Furthermore, since also $\H^{0\dagger} g_{q1}(t)=0$, by virtue of 
\eqref{ApxRRAdjGen}, scalar products of the type
\begin{equation}
\ev{g_{q1}(t)}{\D_1^{0\dagger}\bullet}
=\ev{\D_1^{0\dagger} g_{q1}(t)}{\bullet}
\end{equation}
can be replaced by 
\begin{equation}
\ev{\L^{0\dagger} g_{q1}(t)}{\bullet} = \ev{g_{q1}(t)}{\K\bullet}
\end{equation}
such that \eqref{GKC} becomes
\begin{multline}\label{ApxGKCSimplGK}
L_{\ta\tb} = \Li_{\ta\tb}+\Lr_{\ta\tb} = 
-\frac{1}{\kb\T}\tint \evs{g_{\ta}(t)}{\K g_{\tb}(t)}\\
-\frac{1}{\kb\T}\ttauint \evs{g_{\ta}(t)}
{\K e^{\K\tau}\K g_{\tb}(t-\tau)}
\end{multline}
with $\ta,\tb=wj,q1$.
Next, due to $\K\mathbbm{1}=\L^{0\dagger}\mathbbm{1}=0$, by following 
the same lines, we can replace $g_{\ta}(t)$ with 
$\delta g_{\ta}(t)$ throughout \eqref{ApxGKCSimplGK} thus obtaining
\begin{multline}\label{ApxGKCSimplGK2}
L_{\ta\tb} = 
-\frac{1}{\kb\T}\tint \ev{\delta g_{\ta}(t)}{\K \delta g_{\tb}(t)}\\
-\frac{1}{\kb\T}\ttauint \ev{\delta g_{\ta}(t)}{\K 
\left(\partial_\tau e^{\K\tau}\right)\delta g_{\tb}(t-\tau)}.
\end{multline}
After one integration by parts with respect to $\tau$, this expression
becomes 
\begin{equation}\label{ApxGKCSimplGK3}
L_{\ta\tb}=\frac{(-1)}{\kb\T}\ttauint \ev{\delta g_{\ta}(t)}{
\K e^{\K\tau} \delta \dot{g}_{\tb}(t-\tau)}.
\end{equation}
Here, the upper boundary term vanishes, since the super-operator 
$\K$ is negative semidefinite and the deviations $\delta g_a(t)$ 
are, by construction, orthogonal to its null space, which contains 
only scalar multiples of the unit operator.

An integration by parts with respect to $t$ transforms 
\eqref{ApxGKCSimplGK3} into 
\begin{equation}\label{ApxGKCSimplGK4}
L_{\ta\tb}=\frac{1}{\kb\T}\ttauint \ev{\delta 
\dot{g}_{\ta}(t)}{\K e^{\K\tau} \delta g_{\tb}(t-\tau)},
\end{equation}
where the boundary terms do not contribute due to the periodicity of
the  involved quantities with respect to $t$. 
Finally, another integration by parts with respect to $\tau$ 
yields \eqref{GKCSimplified}.

\section{Reciprocity Relations}\label{ApxRR}

Our aim is to prove the reciprocity relations \eqref{GKCRR}. 
To this end, we have to establish some technical prerequisites. 
First, we introduce the shorthand notation
\begin{multline}\label{ApxRRShorthand}
L_{\alpha\beta}= -\frac{1}{\kb\T}\tint 
\ev{g_\alpha (t)}{\X_{\alpha\beta} g_\beta(t)}\\
-\frac{1}{\kb\T}\ttauint 
\ev{g_\alpha(t)}{\Y_\alpha e^{\K\tau}\Y_\beta g_\beta(t-\tau)},
\end{multline}
where 
\begin{equation}\label{ApxRRDefX}
\left(\!\begin{array}{cc}
\X_{wj,wk} & \X_{wj,q\nu}\\
\X_{q\nu,wj} & \X_{q_\nu, q\mu}
\end{array}\!\right)\equiv
\left(\!\begin{array}{cc}
\K & \D^{0\dagger}\\
\D^{0\dagger} & \delta_{\nu\mu} \D^{0\dagger}
\end{array}\!\right)
\end{equation}
and 
\begin{equation}\label{ApxRRDefY}
\Y_{wj}\equiv \K, \quad \Y_{q\nu}\equiv \D^{0\dagger}.
\end{equation}
Second, we note that \eqref{GKCScalarProd} and \eqref{ApxGKCCummutRel}
imply 
\begin{equation}\label{ApxRRAdjGen}
\ev{\bullet}{\D^{0\dagger}\circ}=\ev{\D^{0\dagger}\bullet}{\circ},
\quad \ev{\bullet}{\K\circ}=\ev{\L^{0\dagger}\bullet}{\circ}.
\end{equation}
Third, by virtue of \eqref{MDHamiltInv}, we have 
\begin{equation}\label{ApxRRGenTRSym}
\D^{0\dagger} =\TR^{-1} \D^{0\dagger}\TR \quad\text{and}\quad
\L^{0\dagger} =\TR^{-1} \K \TR,
\end{equation}
where we used that the time-reversal operator is anti-unitary, i.e.,
$Ti + iT=0$ with $i$ denoting the imaginary unit. 
Combining \eqref{ApxRRAdjGen}, \eqref{ApxRRGenTRSym} with the
definitions \eqref{ApxRRDefX} and \eqref{ApxRRDefY} yields 
\begin{align}
\ev{\bullet}{\X_{\alpha\beta} \circ} 
 & = \ev{\TR^{-1}\X_{\alpha\beta} \TR \bullet}{\circ}
 \quad\text{and}\nonumber\\
\ev{\bullet}{\Y_{\alpha}\circ} 
 & = \ev{\TR^{-1} \Y_{\alpha} \TR \bullet}{\circ}.
\label{ApxRRTRSymRel}
\end{align}
Fourth, from the relation \cite{Mazenko2006}
\begin{equation}
\tr{\bullet} = \tr{\left(T\bullet T^{-1}\right)^\dagger}
\end{equation}
and the fact that $H^0$ commutes with $T$, it follows 
\begin{equation}\label{ApxRRTRSymScalProd}
\ev{\TR^{-1}\bullet}{\circ}=\ev{\bullet^\dagger}{\TR\circ^\dagger}. 
\end{equation}

The reciprocity relations \eqref{GKCRR} can now be obtained through
the calculation
\begin{align}
& L_{\alpha\beta}[g_\alpha(t),g_\beta(t)] 
 = -\frac{1}{\kb\T}\tint \ev{\X_{\alpha\beta}\TR g_\alpha(t)}
   {\TR g_\beta (t)}\nonumber\\
&\quad -\frac{1}{\kb\T}\ttauint
   \ev{\Y_\beta e^{\K\tau}\Y_\alpha \TR g_\alpha (t)}{\TR
   g_\beta(t-\tau)}\nonumber\\[9pt]
& = -\frac{1}{\kb\T}\tint \ev{\TR g_\beta(-t)}
   {\X_{\beta\alpha} \TR g_\alpha (-t)}\nonumber\\
&\quad -\frac{1}{\kb\T}\ttauint\ev{\TR g_\beta(-t)}
  {\Y_\beta e^{\K\tau}\Y_\alpha \TR g_\alpha (\tau-t)}
\nonumber\\[9pt]
&= L_{\beta\alpha}[\TR g_\alpha(-t), \TR g_\beta(-t)].
\label{ApxRRSymProof}
\end{align}
In the first step, we consecutively applied the relations 
\eqref{ApxRRTRSymRel} and \eqref{ApxRRTRSymScalProd} and exploited the
properties $\left(\X_{\alpha\beta}\bullet \right)^\dagger =
\X_{\alpha\beta}\bullet^\dagger$ and $\left(\Y_\alpha\bullet
\right)^\dagger = \Y_\alpha\bullet^\dagger$ of the super-operators
$\X_{\alpha\beta}$ and $\Y_{\alpha}$, which can be
easily found by inspection. 
Furthermore, we used that the operators $g_\alpha(t)$ and $\TR 
g_\alpha(t)$ represent observables and thus must be Hermitian. 
In the second step, we invoked the identities
\begin{align}
& \tint f(t)h(t+\tau) = \tint f(t-\tau) h(t)\quad\text{and}
\nonumber\\
& \tint f(t) = \tint f(\T-t) = \tint f(-t),
\end{align}
which hold for any $\T$-periodic functions $f(t)$ and $h(t)$.
Finally, we used the symmetries $\evs{\bullet}{\circ}=\evs{\circ}
{\bullet}$ and $\X_{\alpha\beta}=\X_{\beta\alpha}$, which are direct
consequences of the definitions \eqref{GKCScalarProd} and 
\eqref{ApxRRDefX}, respectively. 

In the special case, where  
\begin{equation}
g_\alpha(t) = \gamma_\alpha (t) g_\alpha
\end{equation}
with $\gamma_{wj}(t)$, $g_{wj}$ introduced in \eqref{GKCFactCond}, 
$\gamma_{q\nu}(t)$ defined in \eqref{GTDTemp} and $g_{q\nu}\equiv
- H^0$, \eqref{ApxRRShorthand} becomes 
\begin{align}
& L_{\alpha\beta}[\gamma_\alpha(t),\gamma_\beta(t)]=
-\frac{1}{\kb\T}\tint \gamma_\alpha(t)\gamma_\beta(t)
\ev{g_\alpha}{\X_{\alpha\beta} g_\beta}\nonumber\\
& \quad -\frac{1}{\kb\T}\ttauint \gamma_\alpha(t)\gamma_\beta(t-\tau)
\ev{g_\alpha}{\Y_\alpha e^{\K\tau}\Y_\beta g_\beta}
\nonumber\\[9pt]
& = -\frac{1}{\kb\T}\tint \gamma_\alpha(t)\gamma_\beta(t)
\ev{g_\beta}{\X_{\beta\alpha}g_\alpha}\nonumber\\
& \quad -\frac{1}{\kb\T}\ttauint \gamma_\alpha(t)\gamma_\beta(t-\tau)
\ev{g_\beta}{\Y_\beta e^{\K\tau}\Y_\alpha g_\alpha}\nonumber\\
& = L_{\beta\alpha}[\gamma_\beta(t),\gamma_\alpha(t)].
\label{ApxRRSymProofSpec}
\end{align}
Here, for the second identity, we rearranged the scalar products
following the same steps as in \eqref{ApxRRSymProof} and invoked
the condition $\TR g_\alpha = g_\alpha$. 
We thus have proven the relation \eqref{GKCRRFact}.

\section{Role of Quantum Coherence for the Generalized Kinetic Coefficients}\label{ApxQCS}

\subsection{Quasi-Classical Systems}
Our aim is to derive explicit expressions for the quasi-classical kinetic 
coefficients $L_{\alpha\beta}^{{{\rm cl}}}$ introduced in Sec.~\ref{SecClassSys}. 
To this end, we proceed in four steps. 
First, the condition \eqref{GKCCommH} allows us to write the 
perturbations $g_\alpha(t)$ as 
\begin{equation}\label{ApxQCSExpPert}
g_\alpha(t)= \sum_{n=1}^M g_\alpha^n(t) \pro{n},
\end{equation}
where $g_\alpha^n(t)\in\mathbb{R}$ and $\left\{\left|n \right
\rangle\right\}_{n=1}^M$ denotes the set of unperturbed energy 
eigenvectors corresponding to the non-degenerate eigenvalues 
$E^0_1<E^0_2<\cdots <E^0_M$ of $H^0$.
Second, the commutation relations
\begin{equation}
[H^0,V_\nu^\sigma]=\varepsilon^\sigma_\nu V_\nu^\sigma
\quad\text{and}\quad
[H^0,V_\nu^{\sigma\dagger}]=-\varepsilon^\sigma_\nu 
V_\nu^{\sigma\dagger},
\end{equation}
which are part of the detailed balance structure 
\eqref{ApxGKCPropUnpert1}, identify
the unperturbed Lindblad operators $V^\sigma_\nu$ and
$V^{\sigma\dagger}_\nu$ as ladder operators with respect to $H^0$.
Hence, their matrix elements 
with respect to the states $\left|n\right\rangle$ are given by 
\begin{align}\label{ApxQCSLadderOp}
\langle n | V^\sigma_\nu |m\rangle &= 
\Pi(E^0_n-E^0_m-\varepsilon^\sigma_\nu)\langle n 
 V^\sigma_\nu |m\rangle
\quad\text{and}
\nonumber\\
\langle n | V^{\sigma\dagger}_\nu |m\rangle &= 
\Pi(E^0_n-E^0_m+\varepsilon^\sigma_\nu)
\langle n | V^{\sigma\dagger}_\nu |m\rangle
\end{align}
with
\begin{equation}
\Pi(\bullet)\equiv\begin{cases}
1 &\text{for}\quad \bullet=0\\
0 &\text{else}
\end{cases}.
\end{equation}
Third, \eqref{ApxQCSExpPert}, \eqref{ApxQCSLadderOp} and the 
detailed-balance structure \eqref{ApxGKCDBAdjDiss} allow us to rewrite
the expressions \eqref{ApxGKCSolPertLimitCycle} for the first-order 
contributions to the periodic state $\rc(t)$ as 
\begin{align}
\G_{wj}(t)& =\tauint e^{\W\tau}\W\g_{wj}(t-\tau), \nonumber\\
\G_{q\nu}(t)& =\tauint e^{\W\tau}\W_\nu\g_{g\nu}(t-\tau),\nonumber\\
\left\langle n \right| G_\alpha(t) \left| m \right\rangle &= 0
\quad\text{for}\quad n\neq m.
\label{ApxQCSGCorr}
\end{align}
Here, we used the vector notation
\begin{align}
\g_\alpha(t)& \equiv(g_\alpha^1(t),\dots, g_\alpha^M(t))^t,
\nonumber\\
\G_\alpha(t)& \equiv (G_\alpha^1(t),\dots,G_\alpha^M(t))^t
\quad\text{with}\nonumber\\
G_\alpha^n(t)&\equiv\left\langle n \right| G_\alpha(t)
\left| n\right\rangle
\end{align}
and the abbreviation
\begin{equation}
\W\equiv \sum_{\nu=1}^{N_q} \W_\nu,
\end{equation}
where the elements of the matrices $\W_\nu$ are given by 
\begin{equation}\label{ApxQCSDefDetWMatrix}
(\W_\nu)_{mn}\equiv\begin{cases}
\sum_\sigma \Gamma^\sigma_\nu \Pi(E^0_m-E^0_n-
\varepsilon^{\sigma}_\nu) |\langle m|V^\sigma_\nu |n\rangle|^2,
& m>n\\
\sum_\sigma \bar{\Gamma}^\sigma_\nu \Pi(E^0_m-E^0_n+
\varepsilon^{\sigma}_\nu) |\langle n|V^{\sigma}_\nu |m\rangle|^2,
& m<n\\
-\sum_{k\neq m} (\W_\nu)_{mk}, & m=n
\end{cases}. 
\end{equation}
Furthermore the superscript $t$ indicates matrix transposition.
The result \eqref{ApxQCSGCorr} shows that, in first order with respect
to $\Delta_j H$ and $\Delta_\nu T$, the periodic state $\rc(t)$ is 
indeed diagonal in the eigenstates of $H^0$, provided the condition
\eqref{GKCCommH} is fulfilled. 
For the forth step of our derivation, we evaluate
\eqref{ApxGKCWorkCoeff} and \eqref{ApxGKCHeatCoeff}
using \eqref{ApxQCSGCorr} thus obtaining the quasi-classical kinetic 
coefficients 
\begin{widetext}
\begin{align}
L_{wj,wk}^{{{\rm cl}}} &\equiv -\frac{1}{\kb\T}\tint 
\ev{\g_{wj}(t)}{\W\g_{wk}(t)}_{{{\rm cl}}}-\frac{1}{\kb\T}\ttauint 
\ev{\g_{wj}(t)}{\W e^{\W\tau}\W\g_{wk}(t-\tau)}_{{{\rm cl}}},
\nonumber\\ 
L_{wj,q\nu}^{{{\rm cl}}} &\equiv -\frac{1}{\kb\T}\tint 
\ev{\g_{wj}(t)}{\W_\nu\g_{q\nu}(t)}_{{{\rm cl}}}-\frac{1}{\kb\T}\ttauint 
\ev{\g_{wj}(t)}{\W e^{\W\tau}\W_\nu\g_{q\nu}(t-\tau)}_{{{\rm cl}}},
\nonumber\\
L_{q\nu,wj}^{{{\rm cl}}} &\equiv -\frac{1}{\kb\T}\tint 
\ev{\g_{q\nu}(t)}{\W_\nu\g_{wj}(t)}_{{{\rm cl}}}-\frac{1}{\kb\T}\ttauint 
\ev{\g_{q\nu}(t)}{\W_\nu e^{\W\tau}\W\g_{wj}(t-\tau)}_{{{\rm cl}}},
\nonumber\\ 
L_{q\nu,q\mu}^{{{\rm cl}}} &\equiv -\frac{\delta_{\nu\mu}}{\kb\T}\tint 
\ev{\g_{q\nu}(t)}{\W_\nu\g_{q\nu}(t)}_{{{\rm cl}}}-\frac{1}{\kb\T}\ttauint 
\ev{\g_{q\nu}(t)}{\W_\nu e^{\W\tau}\W_\mu\g_{q\mu}(t-\tau)}_{{{\rm cl}}},
\label{ApxQCSGKC}
\end{align}
\end{widetext}
where the simplified scalar product is defined for arbitrary vectors 
$\mathbf{A}\equiv (A_1,\dots,A_M)^t\in\mathbb{R}^M$ and
$\mathbf{B}\equiv (B_1,\dots,B_M)^t\in \mathbb{R}^M$ as 
\begin{equation}
\langle \mathbf{A},\mathbf{B} \rangle_{{{\rm cl}}}
\equiv \mathbf{A}^t\mathbb{P}^{{{\rm eq}}}\mathbf{B}
\end{equation}
with $\mathbb{P}^{{{\rm eq}}}$ denoting the diagonal matrix
\begin{multline}
\mathbb{P}^{{{\rm eq}}}\equiv
{{{\rm diag}}}
(\exp[-E_1/(\kb\Tc)],\\ \dots,\exp[-E_M/(\kb\Tc)])/Z^0.
\end{multline}

The generalized kinetic coefficients \eqref{ApxQCSGKC} describe a 
discrete classical system with periodically modulated energy levels  
\begin{equation}
E_n(t) = E_n^0 + \Delta_j H \sum_{j=1}^{N_{w}} g_{wj}(t),
\end{equation}
whose unperturbed dynamics is governed by the master equation 
\begin{equation}
\partial_t \mathbf{p}(t) = \mathbb{W}^0\mathbf{p}(t).
\end{equation}
Here, the vector $\mathbf{p}(t)\equiv (p_1(t),\dots,p_M(t))^t$ 
contains the probabilities $p_n(t)$ to find the system in the state
$n$ at the time $t$ and the matrix $\mathbb{W}^{0}$ obeys the 
classical detailed balance relation 
\begin{equation}
\mathbb{W}^0\mathbb{P}^{{{\rm eq}}}=
\mathbb{P}^{{{\rm eq}}}\W
\end{equation}
as a consequence of \eqref{ApxGKCPropUnpert1}. 
If $N_q=1$, i.e., if the system is coupled only to a single reservoir,
\eqref{ApxQCSGKC} can be cast into the compact form
\begin{multline}
L_{ab} = -\frac{1}{\kb\T}\tint \ev{\delta\dot{\mathbf{g}}_a(t)}
{\delta\mathbf{g}_b(t)}_{{{\rm cl}}}\\
+\frac{1}{\kb\T}\ttauint \ev{\delta\dot{\mathbf{g}}_a(t)}{
e^{\W\tau}\delta\dot{\mathbf{g}}_b(t-\tau)}_{{{\rm cl}}},
\end{multline}
where $a,b=wj,q1$ and 
\begin{equation}
\delta\mathbf{g}_a(t)\equiv \mathbf{g}_a(t) 
- \mathbf{1}\langle \mathbf{1},\mathbf{g}_a(t)\rangle_{{{\rm cl}}}
\end{equation}
with $\mathbf{1}\equiv (1,\dots,1)^t$. 
These expressions, which here arise as a special case of our general 
result \eqref{GKC}, were recently derived independently in 
\cite{Proesmans2015,Proesmans2015a} by considering a discrete 
classical system from the outset. 

\subsection{Quantum Corrections}

The decomposition \eqref{GKCQCDecomp} can be obtained from the 
following argument. 
First, we note that the super-operator $\H^0$ is skew-Hermitian with
respect to the scalar product \eqref{GKCScalarProd}.
Second, as a consequence of the detailed balance structure
\eqref{ApxGKCPropUnpert1}, the super-operators $\D^{0\dagger}_\nu$
are Hermitian with respect to \eqref{GKCScalarProd} and commute with
$\H^0$.
Consequently, the Liouville space of the system $\mathcal{L}$ can be 
partitioned into subspaces that are orthogonal with respect to 
\eqref{GKCScalarProd} and simultaneously invariant under the action of 
$\H^0$ and each $\D^{0\dagger}_\nu$. 
In particular, such a partitioning is given by the nullsspace of 
$\H^0$, i.e., the set $\mathcal{L}^{{{\rm cl}}}$ of all operators
commuting with $H^0$, and its orthogonal complement $\mathcal{L}^{{{
\rm qu}}}\equiv (\mathcal{L}^{{{\rm cl}}})^{\perp}$.
Since, by construction, $g_{wj}^{{{\rm cl}}}(t)\in\mathcal{L}^{{{\rm
cl}}}$ and $g_{wj}^{{{\rm qu}}}(t)\in\mathcal{L}^{{{\rm qu}}}$, 
\eqref{GKCQCDecomp} now follows directly from the general 
structure of the kinetic coefficients \eqref{GKC}. 

\section{New Constraint}\label{ApxNC}

In order to prove the constraint \eqref{GKCDefA}, we first show 
that the matrix $\mathbb{A}$ defined in is positive semidefinite. 
To this end, we introduce the quadratic form 
\begin{multline}\label{ApxNCDefQ}
\mathcal{Q}(\x,\y,\z)\equiv \x^t\mathbb{L}^{{{\rm ins}}}_{qq}\x
+ 2\x^t \mathbb{L}_{qw}\y + 2\x^t\mathbb{L}_{qq}\z\\
+  \y^t \mathbb{L}_{ww}\y +  \y^t\mathbb{L}_{wq}\z
+  \z^t \mathbb{L}_{qw}\y +  \z^t\mathbb{L}_{qq}\z,
\end{multline}
where $\x\equiv\left(x_1,\dots,x_{N_q}\right)^t, \z\equiv\left(
z_1,\dots,z_{N_q}\right)^t\in\mathbb{R}^{N_q}$ and $\y\equiv\left(
y_1,\dots,y_{N_w}\right)^t\in\mathbb{R}^{N_w}$. 
We will now, one by one, cast the terms showing up on the right-hand
side of \eqref{ApxNCDefQ} into a particularly instructive form. 
To this end, it is convenient to introduce the extended scalar product 
\begin{equation}
\eds{A}{B}\equiv \frac{1}{\T}\tint\evs{A(t)}{B(t)},
\end{equation}
for arbitrary time-dependent operators $A(t)$ and $B(t)$ 

The first term in \eqref{ApxNCDefQ} becomes 
\begin{equation}\label{ApxNCIE1}
\x^t\mathbb{L}_{qq}^{{{\rm ins}}}\x = 
\frac{(-1)}{\kb}\sum_{\nu=1}^{N_q} x_\nu^2
 \ed{g_{q\nu}}{\D^{0\dagger}_\nu g_{q\nu}}.
\end{equation}
after inserting the definition \eqref{GKC} for the coefficients 
$L^{{{\rm ins}}}_{q\nu,q\mu}$. 
Using the expressions \eqref{ApxGKCHeatCoeff}, the second and the 
third one can be respectively written as  
\begin{equation}\label{ApxNCIE2}
2\x^t\mathbb{L}_{qw}\y =\frac{(-2)}{\kb}\sum_{\nu=1}^{N_q}
x_\nu\ed{g_{q\nu}}{\D^{0\dagger}_\nu 
\left(G^y_w + g^y_w\right)}
\end{equation}
and 
\begin{equation}\label{ApxNCIE3}
2\x^t\mathbb{L}_{qq}\z =\frac{(-2)}{\kb}\sum_{\nu=1}^{N_q}
x_\nu\ed{g_{q\nu}}{\D^{0\dagger}_\nu
\left(G^z_q + z_\nu g_{q\nu}\right)}
\end{equation}
with
\begin{equation}
g^y_w(t)\equiv \sum_{j=1}^{N_w} y_j g_{wj}(t), \quad
G^y_w(t)\equiv \sum_{j=1}^{N_w} y_j G_{wj}(t)
\end{equation}
and 
\begin{equation}
G^z_q(t)\equiv \sum_{\nu=1}^{N_q} z_\nu G_{q\nu}(t).
\end{equation}

We now consider the fourth term in \eqref{ApxNCDefQ}. 
By virtue of \eqref{ApxGKCWorkCoeff}, it becomes 
\begin{align}
& \y^t\L_{ww}\y = \frac{(-1)}{\kb}\biggl\{
\ed{g^y_w+G^y_w}{\K\left(G^y_w+ g^y_w\right)}
\nonumber\\
&\hspace*{4cm}
-\ed{G^y_w}{\K\left(G^y_w+g^y_w\right)}\biggr\}
\nonumber\\[9pt]
& = \frac{(-1)}{\kb}\biggl\{
\ed{g^y_w+G^y_w}{\K\left(G^y_w+ g^y_w\right)}
-\ed{G^y_w}{\dot{G}^y_w}\biggr\}\nonumber\\[9pt]
&=\frac{(-1)}{\kb}\sum_{\nu=1}^{N_q}
\ed{g_w^y+G^y_w}{\D^{0\dagger}_\nu
\left(G^y_w+g_w^y\right)}.
\label{ApxNCIE4}
\end{align}
For the second identity, we used the differential equation
\begin{equation}\label{ApxNCDiffEqGw}
\partial_t G^y_w(t) = \K\left(G^y_w(t) + g^y_w(t)\right),
\end{equation}
which derives from \eqref{ApxGKCDiffEq}. 
Since a simple integration by parts with respect to $t$ shows
\begin{equation}\label{ApxNCAntiHermitDt}
\eds{A}{\dot{B}} = -\eds{\dot{A}}{B}
\end{equation}
for arbitrary operatros $A(t)$ and $B(t)$, the contribution
$\eds{G_w^y}{\dot{G}_w^y}=-\eds{G_w^y}{\dot{G}_w^y}$ vanishes.
The third identity in \eqref{ApxNCIE4} then follows by inserting the 
definition \eqref{GKCDefLTR} of $\K$ and noting that $\eds{\bullet}{
\H^0\bullet}=0$ due to 
\begin{equation}\label{ApxNCAntiHermitH}
\eds{\bullet}{\H^0\circ}=-\eds{\H^0\bullet}{
\circ}=-\eds{\circ}{\H^0\bullet}.
\end{equation}
The contributions $\y^t\mathbb{L}_{wq}\z$ and $\z^t\mathbb{L}_{qw}\y$
are most conveniently analyzed together. 
We find 
\begin{widetext}
\begin{align}
\y^t\mathbb{L}_{wq}\z + \z^t\mathbb{L}_{qw}\y 
& = \frac{(-1)}{\kb}
\biggl\{
  \ed{G^y_w + g^y_w}{\K G_q^z
  +\sum_{\nu=1}^{N_q} z_\nu \D^{0\dagger}_\nu g_{q\nu}}
  -\ed{G^y_w}{\K G^z_q + \sum_{\nu=1}^{N_q} z_\nu
   \D^{0\dagger}_\nu g_{q\nu}}\nonumber\\
& \hspace*{3cm}+\sum_{\nu=1}^{N_q}
   \ed{G^z_q+ z_\nu g_{q\nu}}{\D^{0\dagger}_\nu
   \left(G^y_w +g_w^y\right)}
-  \sum_{\nu=1}^{N_q}
   \ed{G^z_q}{\D^{0\dagger}_\nu\left(G^y_w+g^y_w
   \right)}\biggr\}\nonumber\\[9pt]
& = \frac{(-1)}{\kb}
  \biggl\{
  \sum_{\nu=1}^{N_q} \ed{G^y_w + g^y_w}{\D^{0\dagger}_\nu
  \left( G_q^z + z_\nu g_{q\nu}\right)}
  +\ed{G^y_w+ g^y_w}{\H^0 G^z_q}
  -\ed{G^y_w}{\dot{G}^z_q}\nonumber\\
& \hspace*{3cm}+\sum_{\nu=1}^{N_q}
   \ed{G^z_q+ z_\nu g_{q\nu}}{\D^{0\dagger}_\nu
   \left(G^y_w +g_w^y\right)}
  -\ed{G^z_q}{\dot{G}^y_w}
  +\ed{G^z_q}{\H^0\left(G^y_w+g^y_w\right)}
  \biggr\}\nonumber\\[9pt]
& = \frac{(-1)}{\kb}
\sum_{\nu=1}^{N_q}\biggl\{
  \ed{G^y_w + g^y_w}{\D^{0\dagger}_\nu
  \left( G_q^z + z_\nu g_{q\nu}\right)}
 +\ed{G^z_q+ z_\nu g_{q\nu}}{\D^{0\dagger}_\nu
  \left(G^y_w +g_w^y\right)}\biggr\},
\label{ApxNCIE5}
\end{align}
\end{widetext}
where, for the second identity, we inserted the definition 
\eqref{GKCDefLTR} of $\K$ and the differential equations 
\eqref{ApxNCDiffEqGw} and 
\begin{equation}\label{ApxNCDiffEqGq}
\partial_t G^z_q(t) = \K G^z_q(t) + \sum_{\nu=1}^{N_q} z_\nu 
\D^{0\dagger}_\nu g_{q\nu}(t)
\end{equation}
following from \eqref{ApxGKCDiffEq}. 
The third identity in \eqref{ApxNCIE5} is obtained by applying
\eqref{ApxNCAntiHermitDt} and \eqref{ApxNCAntiHermitH}.
Finally, the last term in \eqref{ApxNCDefQ} assumes the form
\begin{align}
& \z^t\mathbb{L}_{qq}\z = \frac{(-1)}{\kb}\sum_{\nu=1}^{N_q}\biggl\{
  \ed{G_q^z +z_\nu g_{q\nu}}{\D^{0\dagger}_\nu\left(
  G_q^z + z_\nu g_{q\nu}\right)}\nonumber\\
&\hspace*{4cm}
  -\ed{G_q^z}{\D^{0\dagger}_\nu\left(
  G_q^z+z_\nu g_{q\nu}\right)}\biggr\}\nonumber\\[9pt]
&= \frac{(-1)}{\kb}\biggl\{\biggl(\sum_{\nu=1}^{N_q}
  \ed{G_q^z +z_\nu g_{q\nu}}{\D^{0\dagger}_\nu\left(
  G_q^z + z_\nu g_{q\nu}\right)}\biggr)\nonumber\\
&\hspace*{4cm}
  -\ed{G^z_q}{\dot{G}^z_q} +\ed{G^z_q}{\H^0 G^z_q}\biggr\}
  \nonumber\\[9pt]
&= \frac{(-1)}{\kb}\sum_{\nu=1}^{N_q}
  \ed{G_q^z +z_\nu g_{q\nu}}{\D^{0\dagger}_\nu\left(
  G_q^z + z_\nu g_{q\nu}\right)},
\label{ApxNCIE6}
\end{align}
where the second identity follows from \eqref{GKCDefLTR} and 
\eqref{ApxNCDiffEqGq} and the third one from \eqref{ApxNCAntiHermitDt}
and \eqref{ApxNCAntiHermitH}. 

Plugging the expressions \eqref{ApxNCIE1}, \eqref{ApxNCIE2}, 
\eqref{ApxNCIE3}, \eqref{ApxNCIE4}, \eqref{ApxNCIE5}, \eqref{ApxNCIE6}
into \eqref{ApxNCDefQ} and recalling \eqref{ApxRRAdjGen} yields 
\begin{equation}
\mathcal{Q}(\x,\y,\z) = -\frac{1}{\kb}\sum_{\nu=1}^{N_q}
\ed{F_\nu}{\D^{0\dagger}_\nu F_\nu}
\end{equation}
with
\begin{equation}
F_\nu(t)\equiv G_w^y(t)+ g_w^y(t) + G_q^z(t) + (z_\nu+x_\nu)
g_{q\nu}(t).
\end{equation}
Since, as a consequence of the detailed balance condition 
\eqref{MDDetailedBalanceLoc}, the super-operators 
$\D^{0\dagger}_\nu$ have only real, non-positive eigenvalues
\cite{Spohn1977,Frigerio1977,Alicki2007}, it follows 
$\mathcal{Q}(\x,\y,\z)\geq 0$ for any $\x,\y,\z$. 
Moreover, the quadratic form \eqref{ApxNCDefQ} can be written as 
\begin{equation}
\mathcal{Q}(\x,\y,\z) = \mathbf{q}^t\mathbb{A}\mathbf{q}
\end{equation}
with $\mathbf{q}\equiv\left(\x^t,\y^t,\z^t\right)^t$ and the matrix
$\mathbb{A}$ defined in \eqref{GKCDefA2}.
We can thus conclude that the matrix $\mathbb{A}$ must be positive
semidefinite. 
The second and the third relation in \eqref{GKCDefA} now follow from
the additive structure \eqref{GKCQCDecomp} of the kinetic coefficients
by setting either $g_{wj}^{{{\rm qu}}}(t)=0$ or 
$g_{wj}^{{{\rm cl}}}(t)=0$.

\section{Quantum Refrigerators}\label{ApxR}

\subsection{Implementation}
In this appendix, we provide a discussion of quantum refrigerators
using the setup and notation of Sec.~\ref{SecQTM}.
To this end, we assume that the thermal gradient $\F_q$ is created 
by two distinct reservoirs with respectively constant temperatures 
$\Tc$ and $\Th>\Tc$. 
The flux $J_q$ then corresponds to the average heat withdrawal from
the hot reservoir in one operation cycle.
Consequently, a proper refrigerator is obtained for
\begin{equation}\label{ApxRCoolingFlux}
\Jc= P- J_q \geq 0.
\end{equation}
Here, $J_q^{{{\rm c}}}$ denotes the heat flux extracted from the cold
reservoir and $-P=\Tc\F_w J_w>0$ the power supplied by the external
controller. 
A common measure for the efficiency of such a device is the 
coefficient of performance \cite{Callen1985}
\begin{equation}
\ve\equiv -\Jc/P\leq \ve_{{{\rm C}}}
\equiv\Tc/(\Th-\Tc),
\end{equation}
where the upper bound $\ve_{{{\rm C}}}$, which corresponds to Carnot
efficiency, follows directly from the second law.
\subsection{Bounds on Efficiency}
Under linear response conditions, the cooling flux
\eqref{ApxRCoolingFlux} becomes 
\begin{equation}\label{ApxRHeatCurr}
\Jc= -(L_{qw}\F_w + L_{qq}\F_q),
\end{equation}
since the power $P$ is of second order in the affinities. 
Together with the expression \eqref{QTMLinFlux} for the work flux
$J_w$, this relation leads to the maximum coefficient of performance
\begin{equation}\label{ApxRMaxCOP}
\ve_{{{\rm max}}}= \ve_{{{\rm C}}}\frac{1}{x}
\frac{\sqrt{1+y}-1}{\sqrt{1+y}+1}
\end{equation}
with respect to $\F_w$ \cite{Benenti2011}.  

In order to show how this figure is restricted by the constraint 
\eqref{GKCDefA}, it is instructive to redefine the
parameter $z$ as 
\begin{equation}
z^{{{\rm c}}}\equiv L^{{{\rm qu}}}_{ww}L_{qq}/L_{qw}^2\geq 0. 
\end{equation}
Relation \eqref{GKCA2CohConstr}, which follows from 
\eqref{GKCDefA}, can then be rewritten as 
\begin{equation}
h_{z}^{{{\rm c}}}\leq y\leq 0 \;\;\text{for}\;\; x<0 
\quad\text{and}\quad 
0\leq y\leq h_{z}^{{{\rm c}}} \;\;\text{for}\;\; x\geq 0
\end{equation}
with $h^{{{\rm c}}}_z\equiv4x/((x-1)^2+4z^{{{\rm c}}})$.
Consequently, we obtain the bound 
\begin{equation}\label{ApxRCOPBnd}
\ve_{{{\rm max}}}\leq \ve_{{{\rm C}}}\frac{1}{x}
\frac{\sqrt{1+h_z^{{{\rm c}}}}-1}{\sqrt{1+h_z^{{{\rm c}}}}-1}
\leq \frac{\ve_{{{\rm C}}}}{1+4z^{{{\rm c}}}}
\end{equation}
with the second inequality being saturated only for $x\rightarrow 0$.
This result proves that cyclic quantum refrigerators, at least in the
linear response regime, can reach Carnot efficiency only in the 
quasi-classical limit, where $L_{ww}^{{{\rm qu}}}=0$ and thus 
$z^{{{\rm c}}}=0$. 
It thus completes our overall picture that coherence effects reduce 
the efficiency of thermal devices. 

We note that the bare current \eqref{ApxRHeatCurr} can not be 
optimized, since it is a unbounded as a function of both affinities. 
Bounding the cooling flux of a refrigerator generally is possible 
only in the nonlinear regime, which is beyond the scope of this 
analysis and will be left to future investigations.

\end{document}